\documentclass[utf8]{FrontiersinHarvard} 
\usepackage{url,hyperref,subcaption}
\usepackage{amsmath}

\newcommand{\asec}{$^{\prime\prime}$}

\def\NH2{$N\rm{(H_2)}$}

\def\SigmaH2{$\Sigma $(${\rm H_2}$)}

\def\AMM{NH$_3$}

\def\METH{CH$_3$OH}

\def\kms{\mbox{km~s$^{-1}$}}
\def\cmc{cm$^{-3}$}

\def\kms{km\,s$^{-1}$}

\def\keyFont{\fontsize{8}{11}\helveticabold }
\def\firstAuthorLast{Fontani} %use et al only if is more than 1 author
\def\Authors{Francesco Fontani\,$^{1,2,3}$}
\def\Address{$^{1}$Istituto Nazionale di Aastrofisica (INAF), Osservatorio Astrofisico di Arcetri, Florence, Italy \\
$^{2}$Max-Planck-Institute for Extraterrestrial Physics (MPE), Garching bei M\"{u}nchen, Germany \\
$^{3}$Laboratoire d'\'Etudes du Rayonnement et de la Mati\`ere en Astrophysique et Atmosph\`eres (LERMA), Observatoire de Paris, Meudon, France}
\def\corrAuthor{Francesco Fontani}

\def\corrEmail{francesco.fontani@inaf.it}

\begin{document}

\onecolumn
\firstpage{1}

\title[P-bearing molecules in ISM]{Observations of phosphorus-bearing molecules in the interstellar medium} 

\author[\firstAuthorLast ]{\Authors} %This field will be automatically populated
\address{} %This field will be automatically populated
\correspondance{} %This field will be automatically populated

\extraAuth{}% If there are more than 1 corresponding author, comment this line and uncomment the next one.
%\extraAuth{corresponding Author2 \\ Laboratory X2, Institute X2, Department X2, Organization X2, Street X2, City X2 , State XX2 (only USA, Canada and Australia), Zip Code2, X2 Country X2, email2@uni2.edu}

\maketitle

\begin{abstract}
 
The chemistry of phosphorus ($^{31}$P) in space is particularly significant due to the key role it plays in biochemistry on Earth.
Utilising radio and infrared spectroscopic observations, several key phosphorus-containing molecules have been detected in interstellar clouds, circumstellar shells, and even extragalactic sources.
Among these, phosphorus nitride (PN) was the first P-bearing molecule detected in space, and still is the species detected in the largest number of sources.
Phosphorus oxide (PO) and phosphine (PH$_3$) were also crucial species due to their role both in chemical networks and in forming biogenic compounds.
The still limited high-angular resolution observations performed so far are shading light on the geometrical distribution of these molecules, which represent crucial insights on their formation processes. 
Observations have also highlighted the challenges and complexities associated with detecting and understanding phosphorus chemistry in space, owing to the low elemental abundance of P relative to other elements.
This review article provides a state-of-art picture of the observational results obtained so far on phosphorus compounds in the interstellar medium. 
Special attention is given to star-forming regions, and to their implications for our understanding of prebiotic chemistry and the potential for life beyond Earth.
Our knowledge of the dominant formation and destruction pathways of the most abundant species has improved, but critical questions remain open, among which: what is (are) the main phosphorus carrier(s) in space?
Upcoming new facilities are expected to contribute significantly to this field, offering opportunities to both detect new phosphorus-bearing molecules and enlarge the number of sources in which the chemistry of P can be studied.
The synergy between observations, theoretical models, laboratory experiments, and computational chemistry is mandatory to significantly progress in our comprehension of the chemistry of this important but poorly studied chemical element. 

\end{abstract}

\section{Introduction}
\label{intro}

The study of interstellar chemistry, that is the chemistry occurring in the interstellar medium (ISM), is an important but difficult task owing to the huge variety of physical conditions and chemical composition in the ISM.
In particular, molecular clouds, that are interstellar clouds where hydrogen is mostly in the form of H$_2$, have densities in the range $\sim 10^2-10^8$~\cmc, and temperatures in the range $\sim 10-100$~K \citep{draine11}.
Stars and planets form in these dense clouds, thus the chemistry occurring in such regions is influenced by the dynamics of the star-formation process, and is known to evolve hand-in-hand with the change in physical properties during the process \citep[e.g.][]{cec12,jorgensen20}.
Having knowledge of such processes is vital to understand the composition of the material out of which stars and planets were born.
Advances in observational techniques have triggered huge progress in this field, thanks to the improved sensitivity, angular resolution, and instantaneous bandwidth coverage of new powerful radio telescopes.
Up to $\sim 240$ different molecules were identified in the ISM until 2022 \citep{mcguire22}, and it is now above 310 (https://cdms.astro.uni-koeln.de/classic/molecules).
The large majority of them contains the most abundant elements in the Universe, namely hydrogen (H), oxygen (O), carbon (C), and nitrogen (N), which are also the most important biogenic elements.
However, two more elements are crucial for pre-biotic chemistry as we know it: sulfur (S) and phosphorus (P).
More than 30 S-bearing molecules were identified in the ISM (https://cdms.astro.uni-koeln.de/classic/molecules), and some of them (e.g. SO, CS, CCS, SO$_2$) are routinely detected in star-forming regions both in the gas-phase \citep[e.g.][]{fuente23,fontani23} and on ice mantles \citep[e.g.][]{boogert15}, as well as in the coma of comets \citep[e.g.][]{calmonte16}.
On the other hand, only seven P-bearing molecules were clearly identified, and one (SiP) tentatively identified.
Table~\ref{tab:molecules} summarises the P-bearing species detected so far in the ISM.
This paucity of identified molecules arises from the relatively low fractional abundance of P with respect to H.
In the present-day Solar photosphere, the [P/H] is $\sim 3 \times 10^{-7}$ \citep{asplund09}.
As reference, that of S is [S/H]$\sim 1.73 \times 10^{-5}$ \citep{lodders03}.

All P-bearing species in Table~\ref{tab:molecules}, except PO$^+$, have been found in the envelopes of evolved stars, while only three (PN, PO, and PO$^+$) were detected in star-forming regions.
However the study of P-compounds is particularly relevant here, due to their high biogenic potential.
Phosphorus is, in fact, a crucial element for the development of life as we know it. 
P-bearing compounds, and in
particular phosphates (PO$^{3-}_4$), are unique in forming large and structurally stable biomolecules, such as deoxyribonucleic acid (DNA) and ribonucleic acid (RNA), phospholipids (the structural components of cellular membranes), and
the adenosine triphosphate (ATP) molecule, which stores the chemical energy within cells.
Therefore, P has an essential role in three of the main features associated to any cellular activity: energy transfer, cell division, and replication \citep[e.g.][]{macia05,schwartz06}.
The search and study of phosphorus molecules in the ISM, and in particular in star-forming regions, is thus essential in the context of astrobiology and our understanding of the emergence of life beyond Earth. 
In particular, several crucial questions, still under debate, are: what is the main source and reservoir of P in space? How do its compounds form and
evolve, and how are they transformed and/or conserved in star-forming regions and, finally, delivered to planets?

The objective of this article is to review the observational studies on P-bearing species in the ISM, and discuss their implications in our current understanding of the astrochemistry of this essential biogenic element.
Emphasis is particularly given to observations towards star-forming regions, and their implications for pre-biotic chemistry.
Even though the focus is on observations, a brief overview on how they fueled theoretical models and laboratory experiments, and vice versa, will be also presented.
The structure of the article is the following: 
Sect.~\ref{origin} briefly discusses the cosmic origin of P, and its measured abundance in the diffuse ISM;
some basic processes in the ISM relevant for P chemistry are presented in Sect.~\ref{basic};
in Sect.~\ref{detection} we give a brief overview of the detection techniques used to identify and analyse P-bearing species;
in Sect.~\ref{sec:molecules}, the observations of the P-bearing molecules identified in the ISM are reviewed;
the input from laboratory experiments and the implications for astrochemical models are discussed in Sects.~\ref{laboratory} and \ref{models}, respectively.
Conclusions and future perspectives are presented in Sect.~\ref{future}.

\begin{table}[]
\caption{Phosphorus-bearing molecules detected in the ISM. \\
$(a)$: Catalog from which $B_{\rm rot}$ and $\mu$ are taken, which is either the Cologne Database for Molecular Spectroscopy (CDMS, https://cdms.astro.uni-koeln.de/cdms/portal/; \citealt{endres16}) or the Jet Propulsion Laboratory \citep[JPL,][]{pickett98}); $(b)$: Tentative detection; $(c)$: from \citet{koelemay22}; $(d)$: parameters relative to the rotation symmetry axis of the molecule. PH$_3$ is a non-linear, oblate symmetric-top rotor. The expression of the energy of the rotational levels is thus different from Eq.(\ref{eq:energy}), but the dependence of $\nu$ of its transitions with $J$ and $I$ is the same. \\ 
References for first detection: 1= \citet{teb87}; 2= \citet{ziurys87}; 3= \citet{tenenbaum07}; 4= \citet{rivilla22}; 5= \citet{guelin90}; 6= \citet{koelemay22}; 7= \citet{agundez07}; 8= \citet{halfen08}; 9= \citet{agundez08}; 10= \citet{tez08}.}
    \centering
    \begin{tabular}{lllccccc}
    \hline
molecule & $B_{\rm rot}$ & $\mu$ & Catalog$^{(a)}$ & First &  Circumstellar & star-forming   & external \\
         & (MHz) & (Dy) & &   detection    &   shells        & regions   & galaxies \\
    \hline
PN         & 23495.2 & 2.75 & CDMS & 1,2 & Y & Y & Y \\
PO         & 21899.5 & 1.88 & CDMS & 3 & Y & Y & N \\
PO$^+$     & 23593.5 & 3.44 & JPL & 4 & N & Y & N \\
CP         & 23859.9 & 0.89 & CDMS & 5 & Y & N & N \\
SiP$^{(b)}$    & 7917$(c)$ & -- & -- & 6 & Y & N & N \\
HCP        & 19976.0 & 0.39 & CDMS & 7 & Y & N & N \\
CCP        & 6372.56 & 3.35 & CDMS & 8 & Y & N & N \\
PH$_3$     & 133480.1$^{(d)}$ & 0.57$^{(d)}$ & CDMS & 9,10 & Y & N & N \\
\hline
    \end{tabular}
%\caption{Phosphorus-bearing molecules detected in the ISM.
%    $(a)$: Catalog from which $B_{\rm rot}$ and $\mu$ are taken, which is either the Cologne Database for Molecular Spectroscopy (CDMS\footnote{https://cdms.astro.uni-koeln.de/cdms/portal/}; \citealt{endres16}) or the Jet Propulsion Laboratory (JPL\footnote{pickett98}); $(b)$: Tentative detection; $(c)$: from \citet{koelemay22}; $(d)$: parameters relative to the rotation symmetry axis of the molecule. 
%    References for first detection: 1= \citet{teb87}; 2= \citet{ziurys87}; 3= \citet{tenenbaum07}; 4= \citet{rivilla22}; 5= \citet{guelin90}; 6= \citet{koelemay22}; 7= \citet{agundez07}; 8= \citet{halfen08}; 9= \citet{agundez08}; 10= \citet{tez08}.   
%    }
\label{tab:molecules}
\end{table}

\section{The cosmic origin of phosphorus and its abundance in diffuse gas}
\label{origin}

Cosmic abundances of all elements are produced through three main mechanisms: Big Bang nucleosynthesis, stellar nucleosynthesis, and neutron capture.
The latter occurs mostly in environments with high fluxes of neutrons, such as in supernova (SN) explosions or in the interior of red giants.
The main phosphorus isotope, $^{31}$P, is traditionally believed to be mainly formed in massive ($M\geq 8 M_{\odot}$) stars by neutron capture on neutron-rich silicon (Si) isotopes, in hydrostatic neon-burning shells of massive star in the pre-SN stage \citep[e.g.][]{clayton03}, and then ejected during SN explosion.
In the explosive phase, irrelevant additional $^{31}$P is created \citep[e.g.][]{wew95,koo13}. 
However, theoretical models in which P is enriched only by core-collapse SNe cannot explain some observed [P/Fe] versus [Fe/H] trends, as derived from spectroscopic observations of Galactic disk stars \citep[e.g.][]{caffau11,roederer14,nandakumar22,maas19}.
\citet{caffau11} also suggested that P production is insensitive to the neutron excess, and hence that processes other than neutron capture should produce it, such as for example proton capture.
A recent study \citep{bet24} is able to explain the anomalous [P/Fe] trends with metallicity if an additional source of P is provided by oxygen–neon (ONe) novae.
In this model, P production occurs at the surface of white dwarfs (WDs) rich in oxygen and neon, whose masses are in between 1.25 and 1.35~$M_{\odot}$.
The progenitor stars of such WDs that evolve in P-rich ONe should have masses $\sim 7-10$~$M_{\odot}$.
To match the observed trends, this would imply a top-heavy initial mass function (IMF), in agreement with observed IMFs measured in low metallicity environments \citep{marks12}.

As previously said (Sect.~\ref{intro}), the present-day Solar photosphere abundance of phosphorus is $\sim 3 \times 10^{-7}$ \citep{asplund09}, in good agreement with meteoritic abundances \citep{lodders03}.
In other Galactic stars similar to the Sun, the P abundance is almost unexplored due to the lack of lines in the commonly accessible wavelength ranges for ground-based telescopes \citep[e.g.][]{caffau11}.
Moreover, the most favourable ultraviolet lines observable are contaminated by stellar continuum \citep{roederer14}.
In the diffuse atomic ISM, P has been detected in the gas-phase in the form of P$^+$ in ultraviolet absorption spectra \citep[e.g.][]{jey78,jenkins86,ses96}.
Its depletion in solid form is a debated topic. 
Like other elements of the third row (e.g. Si and S), it has been for long assumed that it is systematically more depleted than the second row elements in the ISM \citep{turner90}.
In diffuse clouds, \citet{jey78} derived an elemental abundance of $\sim 2 \times 10^{-7}$, indicating a depletion factor of only 2-3.
\citet{dufton86} found no depletion of P in warm diffuse clouds \citep[see also][]{lebouteiller06}, and a depletion of a factor of 3 in cold diffuse clouds, unlike Si which is depleted also in diffuse gas.
\citet{jenkins09} found that the abundance of P in the gas-phase is even super-Solar along some Galactic line of sights \citep[see also][]{ritchey18}.
However, recently \citet{ritchey23} reexamined this result and found P abundances consistent with the Solar one along line of sights where the gas is diffuse and poor in molecules, confirming that P is essentially not depleted here, while increasingly severe
depletion of P is seen along molecule-rich sight lines. 
In the densest and colder molecular clouds of the ISM, the depletion factor can be as high as 600--1000 \citep[e.g.][]{turner90},
although new measurements obtained in massive star-forming regions and evolved star envelopes point to a lower degree of depletion \citep[$\sim 100$][]{rivilla16,ziurys18}.
The chemical models of \citet{chantzos20} predict that significant depletion of P in solid phase can occur only when the volume density of H$_2$ reaches $\sim 10^5$~\cmc, due to freeze-out of atomic P on dust grain mantles at such high densities.

P$^+$ was also detected in the diffuse ISM of external galaxies such as the Large Magellanic Cloud \citep{friedman00}, M33 \citep{lebouteiller06b}, and in more distant low-metallicity star-forming galaxies \citep[e.g.]{lebouteiller13}. 
In particular, the P abundance with respect to O derived by \citet{lebouteiller13} seem in line with P production dominated by massive stars also in such low-metallicity environments.

\section{Basic astrochemical processes involving phosphorus in the ISM}
\label{basic}

Two types of chemical processes are invoked to explain the formation of molecules in the ISM: gas-phase processes and grain-surface processes.
In this section, we give an overview of the gas-phase and grain-surface astrochemical reactions believed to be important for the formation of phosphorus molecules.
Books where these processes are described in detail are, for example, \citet{dew85} and \citet{yamamoto17}.

\subsection{Gas-phase chemistry}
\label{gas}

Gas-phase reactions occur spontaneously if the Gibbs energy, $G$, decreases from reactants to products.
$G$ is a thermodynamical state function, which represents the energy in
isobaric and isothermal reactions, conditions that are normally satisfied in the ISM. 
By definition, $G=H-TS$, where $H=U+PV$ is the enthalpy, $T$ the temperature, $U$ the internal energy, $P$ the pressure, $V$ the volume, and $S$ the entropy of the system.
For an isothermobaric reaction:
\begin{equation}
\Delta G = \Delta U + RT\Delta N - T\Delta S \;,
\label{gibbs}
\end{equation}
where $R$ is the constant of perfect gases and $N$ is the number of moles.
Eq.(\ref{gibbs}) states that at the typical cold temperatures of the molecular ISM ($T \leq 100$~K, corresponding to energies $\leq 0.01$~eV), the terms in $T$ are negligible and hence $\Delta G\sim \Delta U$.
Therefore, in practice only exothermic reactions are spontaneous.
Moreover, some exothermic reactions possess an activation barrier.
Typical activation barriers have energies above 0.01~eV.
Therefore, in practice only exothermic and barrierless reactions are efficient in the dense gaseous ISM.
Among two-body reactions of this kind, (exothermic) ion-neutral reactions are relevant because they do not typically possess activation barriers, owing to the attractive long range electrostatic potential between the ion and the dipole moment induced in the neutral species.
Therefore, if an ion can react with H$_2$, no other reactions need to be considered owing to the much higher abundance ($\geq 10^4$) of H$_2$ with respect to any other neutral.
Among the ion-neutral reactions most relevant to initiate the chemistry of P, in theory that between P$^+$ and H$_2$ would be important.
However, it is endothermic \citep{thorne84} and thus inefficient in molecular gas.
The same is true for the reaction between P$^+$ and CO, the second most abundant neutral molecule.
The first bonds (P-H, P-C, P-O) are thus likely formed by the following proton transfer reactions \citep{millar91}:
\begin{equation}
{\rm P + H_3^+ \rightarrow PH^+ + H_2}\;,
\label{eq:P1}
\end{equation}
\begin{equation}
{\rm P + HCO^+ \rightarrow PH^+ + CO}\;,
\label{eq:P2}
\end{equation}
\begin{equation}
{\rm P^+ + H_2O \rightarrow HPO^+ + H}\;,
\label{eq:P3}
\end{equation}
\begin{equation}
{\rm P^+ + CH_4 \rightarrow PCH_2^+ + H_2}\;,
\label{eq:P4}
\end{equation} 
\begin{equation}
{\rm P^+ + O_2 \rightarrow PO^+ + O}\;.
\label{eq:P5}
\end{equation}
Once PH$^+$ is formed through reactions \ref{eq:P1} and \ref{eq:P2}, binary reactions with H$_2$ and CO cannot occur because endothermic \citep{adams90}, like all reactions between PH$_n^+$ ($n=1-4$) species and H$_2$.
PH$^+$ can however react with H$_2$O to form HPO$^+$, which then, upon dissociative recombination, gives \citep{millar91}:
\begin{equation}
\begin{split}
{\rm HPO^+ + e^-} & {\rm \rightarrow PH + O}\;, \\
                  & {\rm \rightarrow PO + H}\;, \\
                  & {\rm \rightarrow P + O + H}\;.
\end{split}
\label{eq:P6}
\end{equation}

Many neutral-neutral two-body reactions possess activation barriers exceeding $\sim 0.01$~eV, and hence become efficient only in hot ($\geq 100$~K) regions.
However, when one of the reactants is a radical, the activation barrier is usually negligible.
Among neutral-neutral reactions including radicals, relevant for P chemistry, there are \citep[see also][]{jimenez18}:
\begin{equation}
{\rm P + OH \rightarrow PO + H}\;,
\label{eq:P7}
\end{equation}
\begin{equation}
{\rm PH + O \rightarrow PO + H}\;,
\label{eq:P8}
\end{equation}
for PO formation, and:
\begin{equation}
{\rm PO + N \rightarrow PN + O}\;,
\label{eq:P9}
\end{equation}
\begin{equation}
{\rm PH + N \rightarrow PN + H}\;,
\label{eq:P10}
\end{equation}
for PN formation.
The rate coefficients of all these reactions are known only partially.
In the original phosphorus network of \citet{millar91}, only $\sim 30\%$ of reactions had rate coefficients measured in the laboratory.
More measured rate reactions were provided by \citet{anicich93} and \citet{cem94}, the latter work using the rate reactions of \citet{kes83}.
Still, the number of rate reactions well constrained by laboratory measurements is limited.
The existing ones can be taken from the online KInetic Database for Astrochemistry \citep[KIDA][]{wakelam12} and the UMIST Database for Astrochemistry \citep{mcelroy13}.

\subsection{Surface chemistry}
\label{surface}

Surface chemistry, namely the synthesis of molecules on the surfaces of dust grains, 
is extremely important in the ISM because dust grains can help chemical reactions to occur on their surfaces in many ways, playing the role of reactant concentrators, reactant suppliers, chemical catalysts, and third bodies \citep[e.g.][]{del01}.
Surface chemistry was invoked to explain the large amount of H$_2$ in the ISM \citep[e.g.][]{ges63}, as well as the higher-than-expected abundance of some species, in particular complex organic molecules (COMs) in star-forming regions \citep[e.g.][]{ceccarelli23}.
A lot of theoretical and laboratory works have characterised this process in detail \citep[e.g.][]{pirronello99,cet04,wakelam17}.
Grossly, the process can be summarised like this: once a particle hits a dust grain there is a probability, called sticking probability, that
it can be adsorbed on the grain surface.
If the particle is physisorbed, the species is linked to the grain surface via intermolecular forces (like hydrogen bonding, dispersion or electrostatics forces), with binding energy in a range of $0.01-0.2$~eV ($\sim 100-2000$~K).
Light species can move on the surface through thermal hopping or quantum tunneling (Langmuir-Hinshelwood process), and a chemical
reaction can occur if the species encounter reactive partners.
%The (possible) energy released by the reaction is adsorbed by the grain.
If the species are linked to the grain through a chemical bond, with
energy of the order of $\geq 0.1$~eV, the particle cannot move across
the surface.
Thus, a reaction can occur only if
another particle in gas-phase hits the chemisorbed species (Eley-Rideal process).
Once a molecule is created, it can be released in the gas-phase through thermal desorption, cosmic rays or UV rays bombardment, chemical desorption (i.e. desorption due to the energy released by the reaction itself), and grain sputtering.
Photodesorption should be the dominant non-thermal desorption mechanism in diffuse clouds, while cosmic rays bombardment is the dominant one in dense clouds (where UV radiation is shielded), and grain sputtering in shocked regions.

Among surface processes relevant for P chemistry, hydrogenation of atomic P is considered the most relevant \citep{chantzos20} to form sequentially:
\begin{equation}
\begin{split}
    {\rm gH + gP} & {\rm \rightarrow gPH}\;, \\
    {\rm gH + gPH} & {\rm \rightarrow gPH_2}\;, \\
    {\rm gH + gPH_2} & {\rm \rightarrow gPH_3}\;. \\
\end{split}
\end{equation}
The final product, PH$_3$, is in fact predicted to be the main carrier of phosphorus on dust grain surfaces \citep{jimenez18}, and it is expected to desorb at an evaporation temperature of $\sim 100$~K \citep{chantzos20}, or through non-thermal desorption mechanisms as those described above.

\section{Detection techniques}
\label{detection}

\begin{figure}
    \centering
    \includegraphics[angle=0,width=0.9\textwidth]{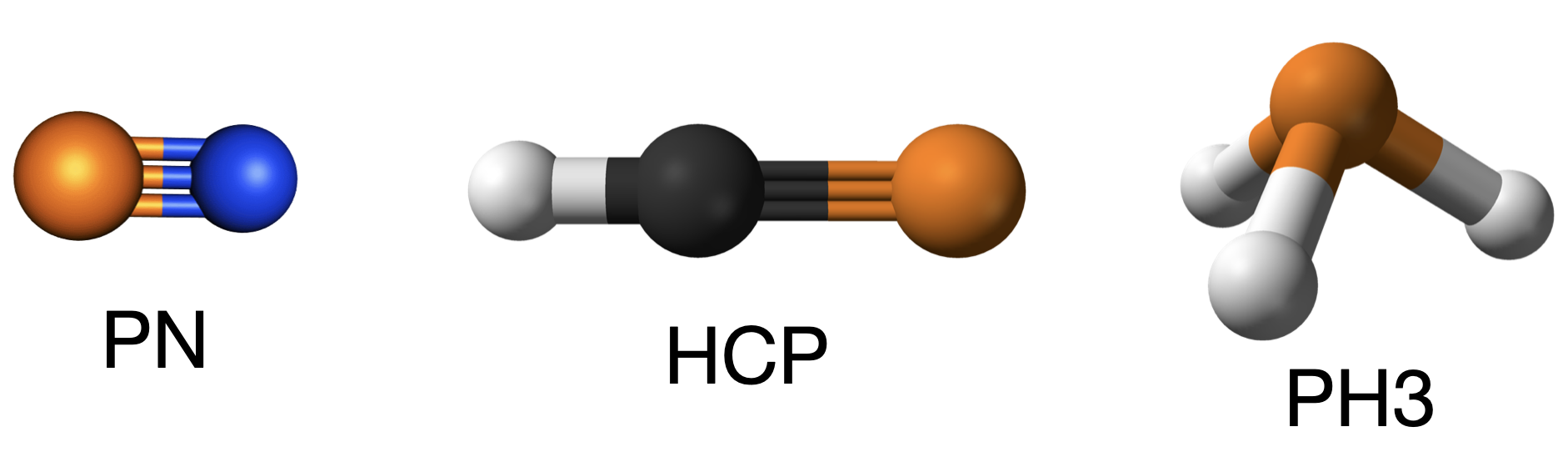}
    \caption{Geometrical structure of PN, HCP, and PH$_3$, as examples of diatomic, triatomic linear, and symmetric-top rotors, respectively.}
    \label{fig:molecules}
\end{figure}
As most of any other molecule \citep[e.g.][]{mcguire22}, the detection of P-bearing species was obtained using radio astronomy techniques in the centimeter and (sub-)millimeter wavelength range.
Molecules can emit and absorb radiation via electronic, vibrational, and rotational transitions \citep[e.g.][]{tes75}.
However, at the typical conditions of the molecular gas in the ISM (see Sect.~\ref{intro}), basically only rotational levels in the ground electronic and vibrational states are populated.
Thus, the molecular emission occurs via transitions between these levels only, whose wavelengths fall in the radio and (sub-)millimeter portion of the electromagnetic spectrum.
This can be shown deriving the energy of the levels of the rotational spectrum, which depends on the geometrical structure of the molecule.
All P-bearing molecules detected so far (Table~\ref{tab:molecules}), except PH$_3$, are linear rotors, in which all nuclei are aligned.
As an example, in Fig.~\ref{fig:molecules}, we show PN and HCP, which are both linear rotors, and PH$_3$ which is a symmetric-top rotor.
For linear rotors, the momentum of inertia of the molecule with respect to the two principal axes perpendicular to the nuclear line, $I$, is the same, and the one along the nuclear line can be neglected.
In such case, one can demonstrate \citep[e.g.][]{tes75} that the energy level of the $J$ rotational state is given, to first order, by the simple equation:
\begin{equation}
E_{\rm J} = hB_{\rm rot} J (J + 1) \;,
\label{eq:energy}
\end{equation}
where $h$ is the Planck constant, and $B_{\rm rot}$ is the rotational constant of the linear rigid rotor, defined as:
\begin{equation}
    B_{\rm rot}=\frac{h}{8\pi^2 I}\;,
\label{eq:brot}
\end{equation}
and it is hence inversely proportional to $I$.
The frequency of an emission transition $J+1\rightarrow J$ is:
\begin{equation}
\nu=2hB_{\rm rot}(J+1) \;, 
\label{eq:nu}
\end{equation}
thus also inversely proportional to $I$.
Because $I$ is larger for heavier molecules and/or molecules with longer structure, molecules containing relatively heavy elements, like P, will have smaller $B_{\rm rot}$ and lower $\nu$ (for the same $J$) with respect to those containing lighter atoms.
Table~\ref{tab:molecules} lists $B_{\rm rot}$ for all P-bearing species detected so far in the ISM.
Putting these $B_{\rm rot}$ in Eq.(\ref{eq:nu}), one finds frequencies for the $J=1-0$ and $J=2-1$ transitions of PN of $\sim 46$~GHz and $\sim 93$~GHz, respectively; for HCP of $\sim 40$~GHz and $\sim 80$~GHz, respectively; and so on.
The energies of the lowest $J$ rotational levels are such that they can be populated even in the coldest portions of the ISM, and emit observable transitions.
This explains why radio-astronomy techniques were, and still are, essential to identify and study P-bearing molecules.

In single-dish radio telescopes, usually the intensity of a molecular transition is expressed in temperature units through the main beam brightness temperature, $T_{\rm MB}$ \citep[see e.g.][]{wilson12}.
The main beam is the main lobe of the diffraction figure of the telescope, in which most of the power received by the antenna falls, and its angular size is considered the angular resolution of the antenna.
If $T_{\rm B}$ is the intrinsic brightness temperature of the source, one can demonstrate that $T_{\rm MB}\sim T_{\rm B}$ for sources more extended than the main beam of the telescope.
On the other hand, if the source is Gaussian and more compact than the main beam, $T_{\rm MB}\sim \sim T_{\rm B}/\eta_{\rm b}$, where $\eta_{\rm b}$
is the so-called "dilution factor".
$\eta_{\rm b}$ is always larger than one, and increases with the decreasing source size.
Therefore, very compact sources will emit lines with observed intensity (i.e. $T_{\rm MB}$) fainter than their intrinsic intensity (i.e. $T_{\rm B}$).
This poses a huge problem, especially for lines of P-bearing molecules which are already intrinsically faint owing to the low elemental abundance of P and depletion in the dense gas.
The need to solve the beam dilution problem, and to improve the angular resolution in images of sources that have a small angular size (compared to the beam size of single-dish telescopes), has led to develop (radio-)interferometers.
As we will see in Sect.~\ref{sec:molecules}, the use of radio interferometers such as the Atacama Large Millimeter Array (ALMA) has allowed to determine the emission morphology of some P-bearing molecules, essential to connect it to the local physical properties.
A general limitation of radio telescopes is that not all molecules emit pure rotational transitions.
In fact, only electric dipole transitions have a sufficient intensity to be detected in a reasonable amount of time, and the strength of these rotational lines is proportional to the square of the permanent electric dipole moment, $\mu$, of the molecule.
This implies that symmetric or almost symmetric rotors, with no or very small $\mu$, can emit only ro-vibrational transitions with appreciable intensity, which, however, occur from levels likely not populated at the typical temperatures of the molecular ISM.
In Table~\ref{tab:molecules} we list $\mu$ of the detected P-bearing molecules.

\section{Observations of phosphorus-bearing molecules in the ISM}
\label{sec:molecules}

The first molecule containing P detected in space is phosphorus nitride, PN, towards three star-forming regions: Ori (KL), W51, and Sgr B2, by observations of its $J = 2-1$, $3-2$, $5-4$, and $6-5$ rotational transitions using the NRAO 12m telescope \citep{teb87} and the FCRAO 14m telescope \citep{ziurys87}.
After a few years, the second P-bearing molecule, CP, was detected in the envelope of the prototypical carbon-rich star IRC+10216 \citep{guelin90}.
For a long time, the detection of phosphorus-bearing molecules remained restricted to these first discoveries.
Advances in instrumental sensitivities allowed in the last $10-15$~yrs to increase significantly both the detection of new P-containing molecules and the number of astronomical sources in which they were detected.
Such studies have triggered new interest in the chemistry of P in the ISM, improving significantly our knowledge of the molecular formation and destruction mechanisms.
In this Section, we give an overview of the observations of P-bearing molecules carried out so far in the two environments where they have been mostly revealed: star-forming regions and evolved stars.
Particular emphasis will be given to the former for their possible implications in pre-biotic processes involving phosphorus.
We conclude the section presenting briefly the detections achieved in other environments, such as extragalactic sources and Solar system objects.

\subsection{Star-forming regions}
\label{starforming}

The first detection of PN in Ori (KL) already suggested that the molecule probably could not be formed in cold gas, because
the measured fractional abundance, [P/H]$\sim 10^{-10}$ \citep{teb87,ziurys87}, was larger by 3-4 orders of magnitude than the expected value if PN was produced by ion-molecule reactions in cold gas.
Moreover, since the P$^+$ and PH$^+$ ions do not react with H$_2$ \citep{thorne84}, a classical ion–molecule reaction scenario did not appear to be efficient to start phosphorus chemistry.
Some laboratory experiments indicated that PO should be the most abundant P-bearing molecule \citep{thorne84}.
However, the non detection of PO in Ori (KL), Sgr B2, and DR21 (OH) \citep{matthews87}, all detected in PN, conducted \citet{teb87} to the conclusion that PN is produced by processes that do not also produce PO (as well as HCP and PH$_3$).
Among these processes, grain disruption was proposed to be the most promising one, forming PN as indirect product via atomic P reacting with, for example, NH$_3$, known to be efficiently formed on dust grains and then released in the gas-phase upon grain sputtering.
However, theoretical works proposed that a gas-phase reaction network could be able to reproduce the observed abundances without the need of shocks, and predict that PN should be the only species likely to be detected if P depletion is larger than 100.
For example, the model of \citet{millar87} predict that, if H$_2$O is abundant, PN is efficiently formed in the gas from
reactions \ref{eq:P3} followed sequentially by reactions \ref{eq:P6}, \ref{eq:P9}, and \ref{eq:P10}.
%\begin{equation}
%{\rm P^+ + H_2O \rightarrow HPO^+ + H} \;,
%\end{equation} 
%followed by:
%\begin{equation}
%{\rm HPO^+ + e^- \rightarrow PH + O} \;, 
%\end{equation}
%\begin{equation}
%{\rm HPO^+ + e^- \rightarrow PO + H} \;,
%\end{equation}
%\begin{equation}
%{\rm PH + N \rightarrow PN + H} \;,
%\end{equation}
%\begin{equation}
%{\rm PO + N \rightarrow PN + O} \;.
%\end{equation} 
Similarly, \citet{cem94} suggested that in hot cores PN can be efficiently formed in hot gas upon evaporation from dust grains and subsequent destruction of PH$_3$, rapidly (in less than $10^4$ yrs) converted into P, PO, and PN.
Nevertheless, a search for P-bearing species (HCP, HPO, and PH$_3$) in larger samples \citep{turner90} gave a few more PN detections in hot regions (2 clear and 1 tentative), but still no detections in cold cores, nor in other P-bearing species in any target.
Clearly, a significant progress in the theoretical debate could not take place with such low number of observational constraints.

After about 20 years from the first discoveries, in 2011 \citet{yamaguchi11} reported the first detection in a low-mass star-forming region of a phosphorus molecule: PN in the $J=2-1$ line towards L1157 B1 and B2, two well-studied chemically rich protostellar bow shocks formed by the interaction between the ambient material and the outflow from the low-mass protostar IRAS 20386+6751 \citep[e.g.][]{teb95}.
The derived [PN/H$_2$] fractional abundances in B1 and B2 are of the order of $10^{-10}$.
The non-detection towards the protostar indicated again that the PN emission is associated preferentially with shocked material.
\citet{fontani16} have doubled the number of star-forming regions detected in PN through observations with the IRAM 30m telescope towards a sample of high-mass star-forming regions in different evolutionary stages, from starless cores to ultracompact HII (UCHII) regions.
They reported the detection of PN $J=2-1$ in 2 starless cores, 3 protostellar objects, and 3 UCHIIs, bringing the number of firmly detected sources in star-forming regions to 14 (13 high-mass and one low-mass).
The observations of \citet{fontani16} also report for the first time PN detections in relatively quiescent material. 
In fact, the measured line widths at half maximum were below $\sim 2$~\kms\ in two targets.
This brought into consideration again the theoretical possibility to form PN also in not shocked gas.

\begin{figure*}
    \centering
    \includegraphics[width=15.5cm]{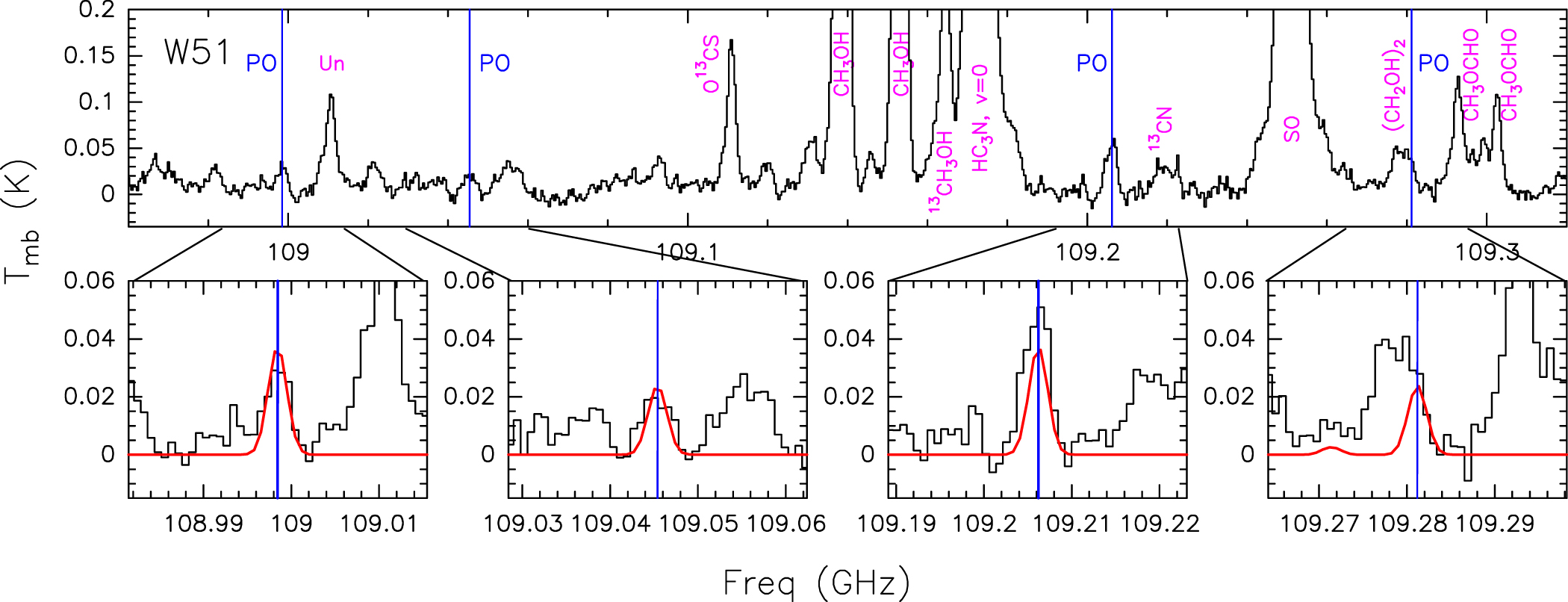}
    \includegraphics[width=15.5cm]{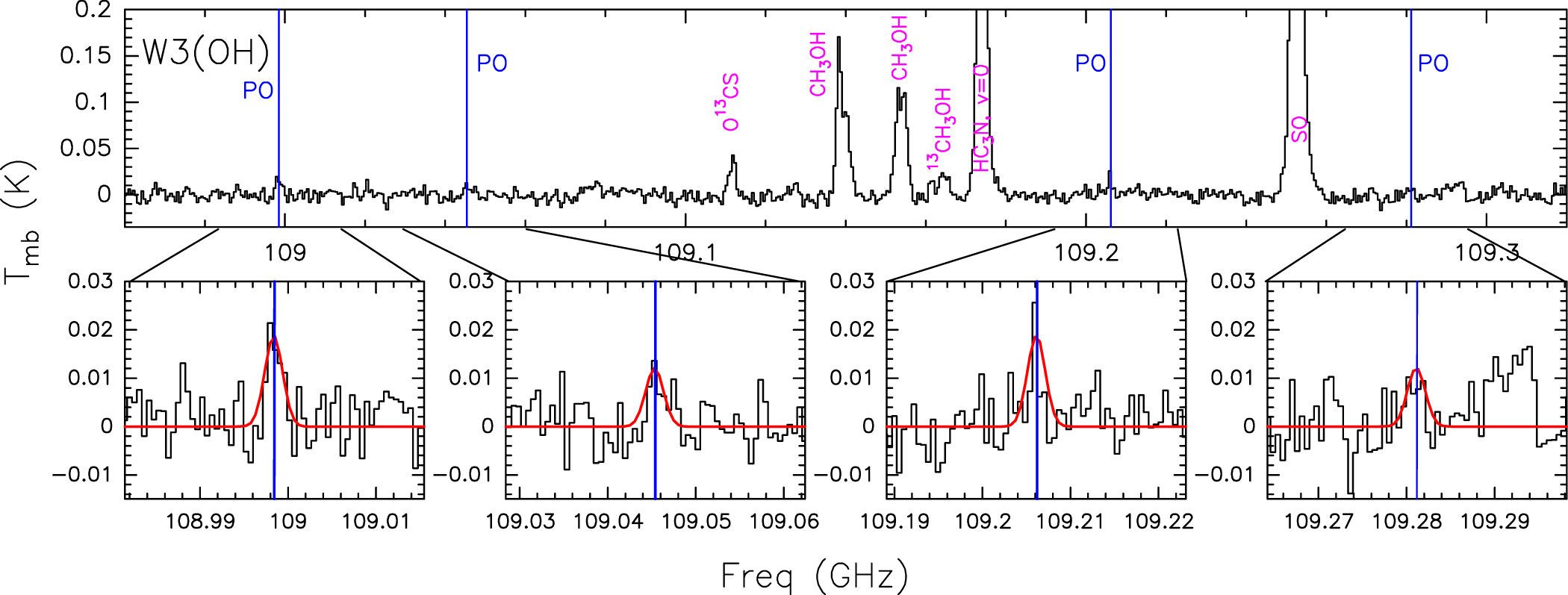}
    \caption{{\it Top panels:} Spectrum observed at 3 mm toward W51. 
    The PO transitions are indicated with blue vertical lines. 
    Under the spectrum, zoom-in views of the PO transitions are highlighted. 
    The red line is the fit in local thermodynamic equilibrium with an excitation temperature of 35~K and an assumed source size of 12\asec.
    {\it Bottom panels:} same as top panels for W3(OH).
    From \citet{rivilla16}.}
    \label{fig:PO}
\end{figure*}

In the same year as the PN survey of \citet{fontani16}, \citet{rivilla16} reported the first detection of phosphorus oxide (PO) in two star-forming regions, W51 and W3(OH) (both luminous high-mass objects), followed shortly after by \citet{lefloch16} towards L1157 B1.
Such new detections were particularly important for many reasons. 
First, PO was predicted to be the most abundant P-bearing species in laboratory experiments \citep{thorne84}.
Second, it is the basic bond of phosphates, and hence understanding its formation and survival in the ISM is crucial for astrobiology.
Figure~\ref{fig:PO} shows the lines of PO detected in W51 and W3(OH) by \citet{rivilla16}.
The abundance ratio PO/PN measured by \citet{rivilla16} and \citet{lefloch16} is $\sim 1.8-3$ and $\sim 2.8$, respectively, indicating a PO abundance higher than PN by a comparable amount.
To understand the origin of PN and PO emission, \citet{rivilla16} used the \citet{veh13} model which simulates the chemical evolution of a parcel of gas and dust with time-dependent physical conditions: 
a cold collapsing core is followed by a warm-up phase simulating a newborn protostar.
The model indicates that the two molecules are formed via gas-phase ion–molecule and neutral–neutral reactions during cold collapse, they freeze-out on dust grains during the early cold phase, and then are both released into the gas when the temperature reaches $\sim 35$~K during core warm-up. This scenario needs two conditions to be satisfied: (1) the P elemental abundance should be [P/H]$\sim 5 \times 10^{-9}$; (2) the gas temperature should be in between $\sim 35$~K and $\sim 90$~K.
The first condition, in particular, indicates a P abundance 25 times higher than typically assumed ($\sim 2 \times 10^{-10}$), or, said in other words, that P should be less depleted than usually assumed.
\citet{lefloch16} used the shock model in \citet{viti11} to reproduce the PN and PO emission towards L1157 B1.
Again, the total P abundance in the gas-phase should be of the order of $10^{-9}$ to match the observations.
\citet{lefloch16} suggest that in L1157 B1 PN and PO are both produced in the shock upon release of PH$_2$ from dust grains, with PO produced at later times than PN.
Moreover, the key player regulating the relative abundance PO/PN is atomic nitrogen, the abundance of which depends on the shock velocity that regulates the conversion of N to \AMM, and vice versa.
In summary, the two studies agree in the need for a gas-phase abundance of P higher than expected, but while \citet{rivilla16} indicate that PN and PO are a {\it direct} product of grain mantle desorption, \citet{lefloch16} indicate an {\it indirect} production in gas-phase upon release of PH$_2$ from the grains.

\begin{figure}[h]
\begin{center}
\includegraphics[angle=0,width=0.95\textwidth]{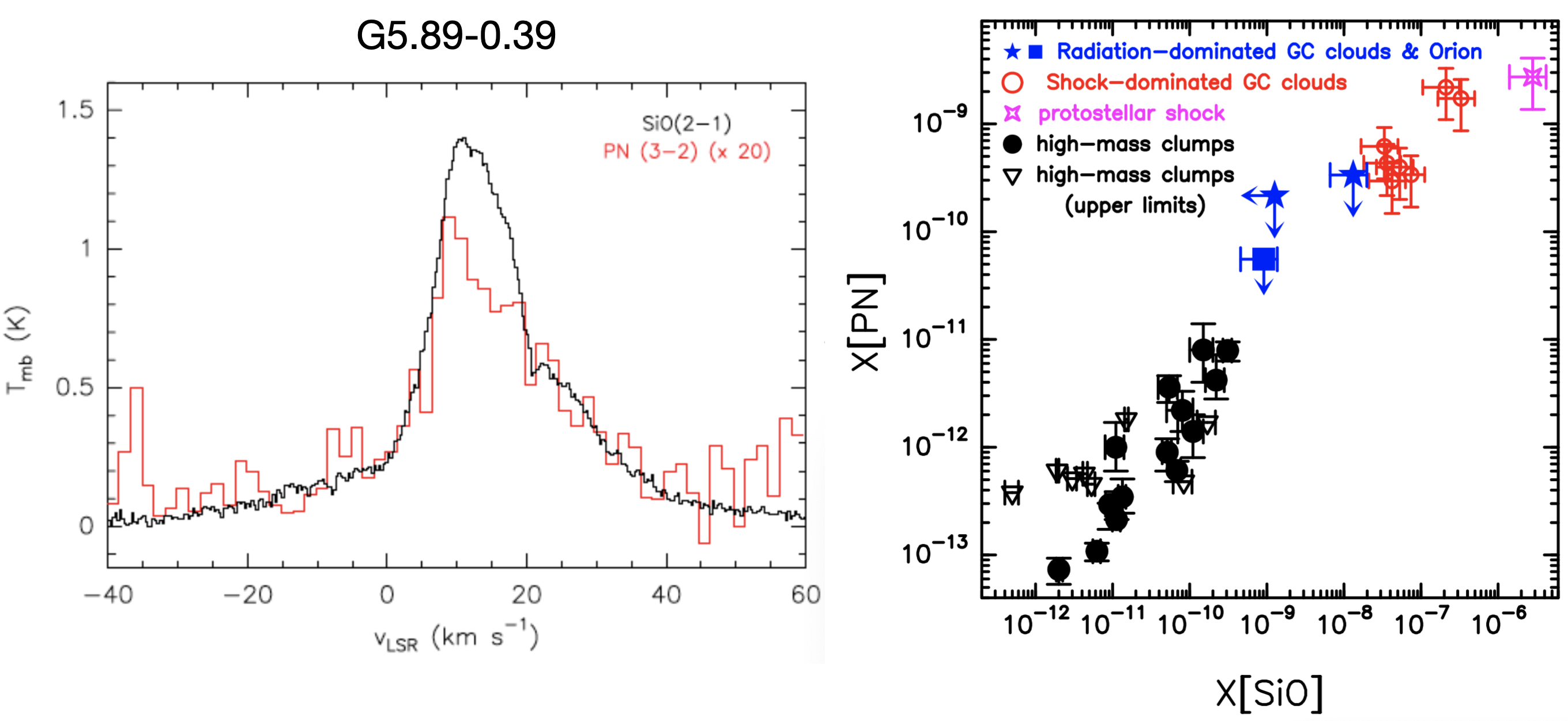}
\end{center}
\caption{{\it Left panel:} Spectrum of PN (3–2) (red histogram; multiplied by a factor of 20) and SiO (2–1) (black histogram) lines measured towards the G5.89--0.39 UCHII region. Taken from \citet{mininni18}.
{\it Right panel:} Fractional abundance of PN, $X$[PN], against that of SiO, $X$[SiO], in different Galactic star formation environments:
High-mass star-forming clumps \citep[black filled circles,][upper limits are indicated by empty triangles]{fontani19}, shock-dominated Galactic Centre clouds \citep[red circles][]{rivilla18}; radiation-dominated Galactic Centre clouds \citep[blue stars][]{rivilla18}, the Orion Bar (blue square; Cuadrado,
private communication), the L1157 B1 shock \citet[magenta open star,][]{lefloch16}. 
The $X$[SiO] in \citet{rivilla18} were computed converting the $^{29}$SiO column density in $^{28}$SiO column density assuming an isotopic ratio $^{28}$Si/$^{29}$Si=19.6 \citep{aeg89}. 
Moreover, the abundances of both PN and SiO were converted from PN/$^{34}$CS and SiO/$^{34}$CS to PN/H$_2$ and SiO/H$_2$ assuming H$_2$/$^{34}$CS$\sim 8.8\times 10^{11}$ \citet{wilson99}.
Adapted from \citet{fontani19}.}
\label{fig:corr-PN-SiO}
\end{figure}

\citet{rivilla18} shed more light on the role of shocks in the formation and destruction of PN and PO observing seven star-forming regions in the Galactic Centre, characterised by different types of chemistry.
They detected five out of seven regions in PN and only one in PO, and suggested an efficient formation of these molecules in shock-dominated regions upon grain sputtering, and an efficient destruction in radiation-dominated (UV/X-rays/cosmic ray) regions.
Such photo-destruction would change the PO/PN ratio, since PO is predicted to be destroyed more efficiently than PN by UV photons \citep{jimenez18}.
We will come back to this point when discussing the extragalactic detection of PN (Sect.~\ref{galaxies}).
\citet{mininni18} and \citet{fontani19} followed-up the PN $J=2-1$ line survey of \citet{fontani16} again with single-dish telescopes but in an enlarged sample and utilising multiple lines.
Both studies agree that the derived excitation temperatures of PN are in the range $5-30$~K, and are likely sub-thermally excited.
Another result that the two works have in common is that the PN line profiles resemble those of SiO, a well-known shock tracer (see left panel in Fig.~\ref{fig:corr-PN-SiO}).
In particular, \citet{fontani19} found that PN is detected only in sources having non-Gaussian high-velocity wings in the lines of SiO, and confirmed the positive correlation between the PN and SiO abundances (see right panel in Fig.~\ref{fig:corr-PN-SiO}) previously suggested in the Galactic Centre clouds by \citet{rivilla18}.
All these results converge to the same conclusion: shocks are necessary to form PN no matter if they are a direct or a secondary product of dust grain mantles.
The PN emission detected in relatively quiescent gas by \citet{fontani16} could thus be originated in shocked material which had sufficient time to cool down and become more quiescent.

From the side of low-mass star formation regions, \citet{bergner19} detected PN and PO in multiple lines for the first time in the envelope of a low-mass protostar: B1-a.
The study was then followed-up in seven similar protostars, all of them known to be associated with an outflow \citep{web22}.
%Such class I protostar is associated with an outflow mapped in SiO \citet{bergner19}.
The multi-line analysis performed in these studies indicates that also in the low-mass regime: 
(1) both the PN and the PO emission are likely sub-thermally excited; 
(2) the line profiles of PN and PO resemble those of the shock tracers SiO, SO$_2$, and \METH; 
(3) the PO/PN relative abundance ratio is in the range 0.6--2.2, confirming that PO tends to be more abundant than PN.
The latter result, in particular, is in common with the other objects in which both molecules were detected, all characterised by the presence of shocked gas but associated with different physical properties: a protostellar bow-shock, two very luminous high-mass star-forming regions, and a Galactic Centre molecular cloud.

\begin{figure}[h]
    \centering
    \includegraphics[width=0.9\textwidth]{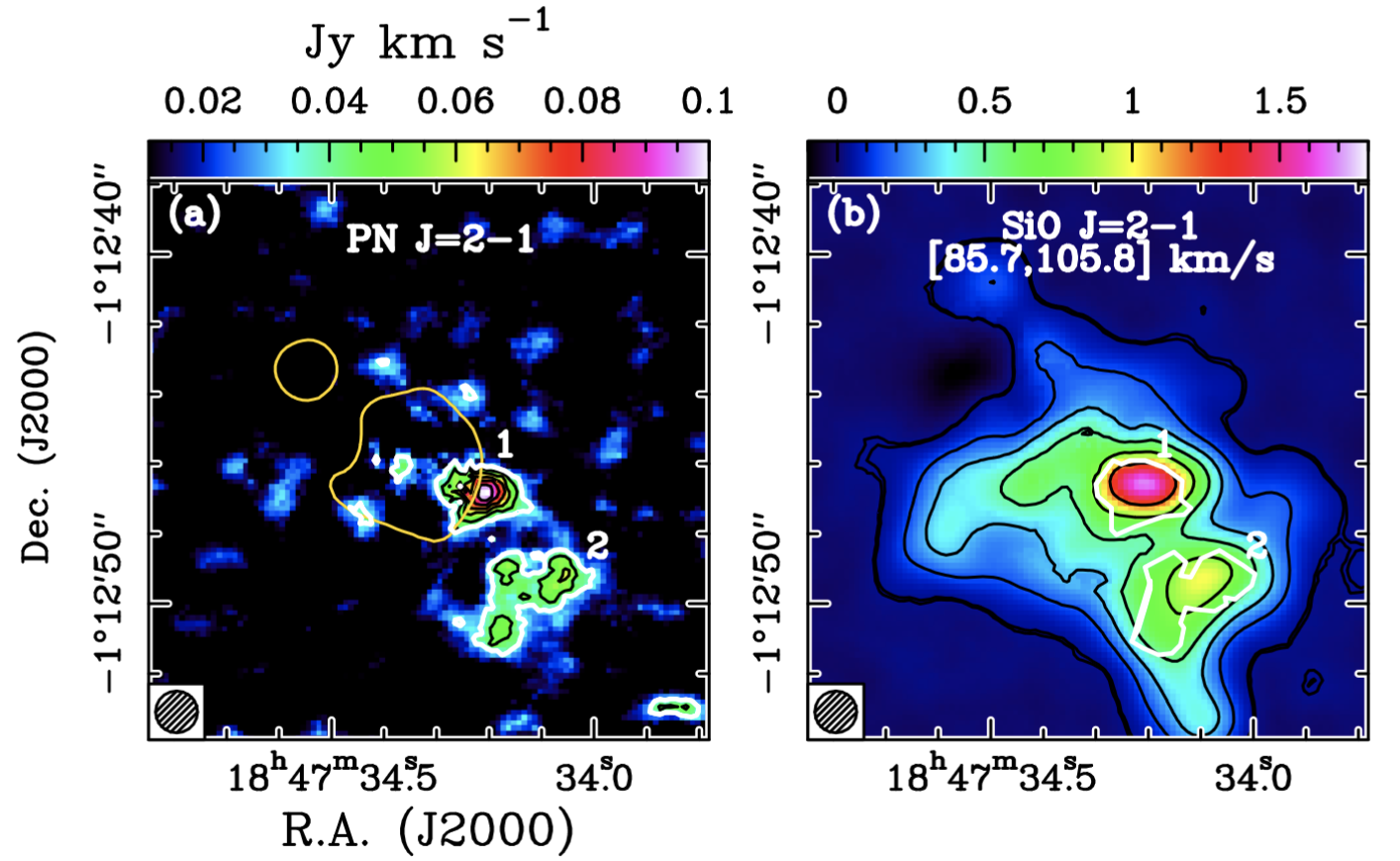}
    \caption{Intensity maps of PN and SiO integrated in velocity towards G31.41+0.31 \citep{fontani24}, obtained with ALMA.
    (a) PN $J=2-1$ integrated in the range 85.7--105.8~\kms. 
    The white contour is the $3\sigma$ rms level of the integrated map ($\sigma = 1.14\times 10^{-2}$ Jy \kms), while the black contours are in steps of $1\sigma$ rms. The PN emission arises from two regions labelled as 1 and 2, offset from the yellow contour which is the 3~mm continuum emission ($20\sigma$~rms) from the hot core. 
    The synthesised beam is in the bottom-left corner. 
    (b) Map of the intensity of SiO $J=2-1$ integrated in the same velocity range as PN (colour scale). 
    The PN emission regions identified in panel (a) are highlighted in white. 
    Contours start at the $3\sigma$ rms level of the integrated emission ($3\times 10^{-2}$ Jy \kms), and correspond to 3, 15, 30, 50, 80, and 120$\sigma$.
    Adapted from \citet{fontani24}.}
    \label{fig:G31}
\end{figure}

The tight association between PN, PO, and shock emission was robustly confirmed by high-angular resolution studies.
\citet{rivilla20}, \citet{bergner22} and \citet{fontani24} mapped at high-angular resolution the PN and PO emission towards three protostars with different properties: the high-mass protostellar object AFGL 5142 \citep{rivilla20}, the low-mass class I protostar B1-a \citep{bergner22}, and the prototypical chemically rich hot core G31.41+0.31 \citep{fontani24}.
All sources are driving outflows associated with typical shock tracers: SiO, SO, and SO$_2$.
Again, despite the different instrinsic properties of the targets (luminosity, mass, evolutionary stage), the results are similar.
First, the phosphorus molecules emit from compact spots which coincide with regions where the protostellar outflows interact with environmental dense and quiescent gas. 
As an example, Fig.~\ref{fig:G31} shows the emission morphology of PN and SiO in the G31.41+0.31 star-forming region: the similar spatial distribution of PN and SiO emission is apparent, even if PN emission is more compact.
Second, PN and PO emission is generally cospatial with low-velocity and not with high-velocity SiO and SO emission.
This could be due to either different solid carriers of P and Si/S in the grains, or to insufficient sensitivity in the high-velocity range of the PN and PO spectra.
In any case, the lack of a clear detection of PN towards the protostellar envelopes of AFGL 5142, B1-a, and G31.41+0.31 \citep[][respectively]{rivilla20,bergner22,fontani24}, traced by the 3~mm continuum in Fig.~\ref{fig:G31}, allows one to rule out relevant formation pathways in hot gas.

The third (and last so far) P-bearing molecule detected in a star-forming regions, PO$^+$, was found by \citet{rivilla22} towards the Galactic Centre molecular cloud G+0.693-0.027.
This is the first phosphorus molecular ion found in the ISM, which had hence important implications in the ionisation efficiency of P-bearing species.
Comparing the abundance of PO$^+$ with those of other similar ions (SO$^+$ and NO$^+$) and to the predictions of chemical models, \citet{rivilla22} concluded that a very high cosmic ray ionisation rate, $\zeta$, is needed ($\zeta\sim 10^{-15}-10^{-14}$ s$^{-1}$) to explain the observed [PO$^+$/H$_2$] of $\sim 4.5 \times 10^{-12}$.
Such high $\zeta$ is indeed measured in G+0.693-0.027.
The main source of PO$^+$ should be atomic P, produced from dissociation of PH$_3$ (desorbed from dust grain mantles), and then ionised by a large flux of cosmic rays.
PO$^+$ is then formed by P$^+$ + O$_2$ and/or P$^+$ + OH.
An alternative, or complementary, formation pathway for PO$^+$ could be direct ionisation of PO, which in any case needs a high $\zeta$ to explain the high observed PO$^+$/PO column density ratio.
Rate coefficients for radiative association of P$^+$ and O were computed by \citet{qin23}, but it is unclear how relevant this process can be in producing PO$^+$. 

A key ingredient in phosphorus chemistry is certainly phosphine, PH$_3$.
Phosphine is formed via hydrogenation of H on dust grains \citep[because gas-phase routes are endothermic and hence inefficient,][]{thorne84}, and then desorbed via thermal \citep[][]{chantzos20,mek21} or non-thermal \citep[e.g.][]{jimenez18,furuya22} processes.
So far, phosphine has never been detected in a star-forming region, indicating either inefficient desorption mechanisms, or a rapid destruction and conversion into other P-bearing compounds \citep[e.g.][]{jimenez18}.
In the prototypical pre-stellar core L1544, a stringent upper limit abundance of $\sim 5 \times 10^{-12}$ in the central part of the core was derived by \citet{fes24} with ALMA. 
Based on their astrochemical modeling of the source, they infer an upper limit to the volatile
P elemental abundance of $5\times 10^{-9}$, 
%in agreement with the values inferred by \citet{rivilla20},
consistent once again with a depletion level in dense gas lower than typically assumed.
Towards the chemically rich protostellar bow-shock L1157 B1, \citet{lefloch16} derived a PH$_3$ upper limit abundance of $\sim 10^{-9}$.
Therefore, both PN and PO are more abundant than PH$_3$ here, consistent with a rapid conversion of PH$_3$ into these, and maybe other, species after evaporation from grain mantles.

Finally, although phosphorus is thought to be mainly produced in massive stars and injected in the environment through SN explosions (see Sect.~\ref{origin}), P-bearing molecules were detected also in WB89-621, a star-forming region located in the outskirt of the Galactic disk \citep{koelemay23} where the presence of supernovae is extremely rare.
WB89-621 is at a Galactocentric distance of 22.6~kpc \citep{blair08}, even though a more recent estimate place it at $\sim 18.9$~kpc \citep{fontani22}.
In any case, the source is at a distance from the Galactic Centre where metallicity is expected to be $\sim 4-5$~times lower than Solar \citep[e.g.][]{shimonishi21}.
The measured PN and PO abundances, comparable with those in the local Galaxy \citep{koelemay23} in an environment where supernovae are basically absent \citep{ranasinghe22}, supports the idea that massive stars and supernovae are not the unique sources of phosphorus (see Sect.~\ref{origin}).

\subsection{Evolved stars}
\label{evolved}

Even though this review is mostly focussed on star-forming regions, a significant contribution on the interstellar chemistry of phosphorus was given by evolved stars.
As said in Sect.~\ref{intro}, all phosphorus molecules found in the ISM except PN and PO$^+$ were detected for the first time in circumstellar shells of evolved stars, in particular carbon and oxygen rich stars.
The CP radical was the first phosphorus molecule detected towards IRC+10216 \citep{guelin90}, a carbon-rich star very prolific in providing detections of new molecules in space \citep[e.g.][]{mcguire22}.
The most obvious processes invoked to explain CP formation, namely photodissociation of HCP or ion-molecule reactions in the outer part of the envelope \citep{glassgold87}, were both challenged by
the fact that the emission seems confined in an inner shell.

\begin{figure}
    \centering
    \includegraphics[width=1\textwidth]{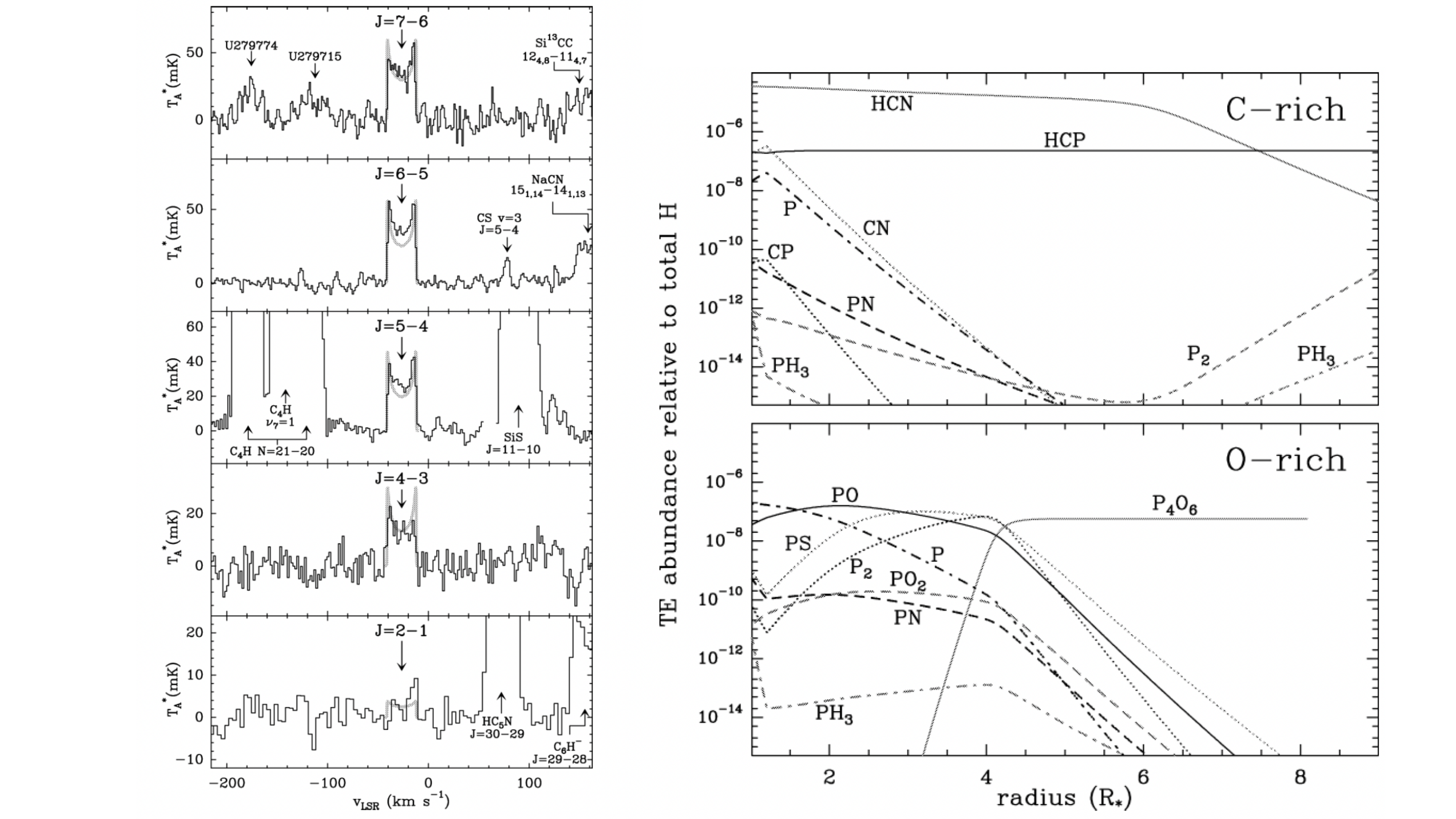}
    \caption{{\it Left panel}: rotational transitions $J=2-1$, $4-3$, $5-4$, $6-5$, and $7-6$ (from bottom to top) of HCP observed toward IRC+10216 with the IRAM 30 m telescope by \citet{agundez07}. 
    Thick gray lines show the line profiles predicted from a radiative transfer model and the HCP abundance profile in the right panel.
    {\it Right panel}: Thermochemical equilibrium abundances predicted in the innermost region of a circumstellar envelope (C‐rich: upper panel; O‐rich: lower panel) as a function of radius from the star.
    Taken from \citet{agundez07}.}
    \label{fig:agundez}
\end{figure}

For ten years, CP remained the only phosphorus species found in a circumstellar shell, until \citet{cernicharo00} detected PN in the same object.
Then, a boost of new detections of phosphorus molecules happened in 2007 and 2008.
Again towards IRC+10216, the third P-bearing species, HCP, was detected in multiple lines \citet{agundez07}.
They are shown in the left panel of Fig.~\ref{fig:agundez}.
The line profiles indicated again that this species is confined to the inner envelope, probably near the stellar photosphere.
As discussed in \citet{agundez07}, the gas chemical composition of the atmosphere of cool stars like IRC+10216 is in thermal equilibrium (TE).
The prediction of TE models is that up to 2-3 stellar radii the dominant gaseous P-bearing molecule depends on the C/O relative ratio.
These predictions are shown in the right panel of Fig.~\ref{fig:agundez}.
In C-rich stars like IRC+10216, all gaseous P is in the form of HCP, which should have an abundance more than three orders of magnitude higher than any other phosphorus molecule.
However, the observed HCP line intensities suggest already a significant depletion of HCP (due to either freeze-out or dissociation) already at a few stellar radii.
That the emission of phosphorus molecules is associated with inner shells of circumstellar envelopes was further confirmed by the first PO detection, the fourth P-bearing molecule seen in the ISM, reported by \citet{tenenbaum07} toward the oxygen-rich supergiant star VY Canis Majoris (VY CMa).
The line profiles of the detected PO and PN transitions suggest an origin again in the inner part of a spherical wind known to be present in VY CMa \citep{ziurys07}.
The same conclusion was drawn by observations of \citet{milam08} towards the C-rich protoplanetary nebula CRL 2688, who found that HCP and PN should be produced in the inner shells, and CP is likely a product of HCP photodissociation at larger radii.
However, \citet{tenenbaum07} measured comparable abundances of PN and PO in VY CMa, which is at odds with the prediction of circumstellar chemistry models for O-rich stars: considering only processes in TE, these models predict a PO abundance much higher ($\sim 2$ orders of magnitude) than that of PN.
Therefore, \citet{tenenbaum07} invoked non-equilibrium shock processes, as proposed by \citet{cherchneff06}, to explain the comparable PN and PO abundances in VY CMa \citep[see also][]{milam08}.

More recently, new identification of PO and PN towards the circumstellar shells of several O-rich stars were reported: the Asymptotic Giant Branch (AGB) stars TX Cam and R Cas \citep{ziurys18}, the supergiant NML Gyg \citep{ziurys18}, and the Mira-type variable star IK Tau \citep{debeck13}.
Both the observed compact emission of PN and PO \citep{debeck13}, and the predictions of Non-TE models applied to a spherical radial distribution of the gas, suggest that the two molecules are always formed near the stellar photosphere, perhaps with their abundances enhanced by shocks \citep{ziurys18}.
Moreover, the observed molecular abundances indicate gas phase carriers of P even more abundant than previously thought. 
Another important piece of the puzzle was added by the detection of the fifth and sixth phosphorus molecules in the ISM: CCP and PH$_3$.
The CCP radical, detected by \citet{halfen08} towards IRC+10216 in four rotational transitions, shows line profiles and abundance again consistent with an inner shell production, perhaps from radical-radical reactions involving CP, or ion-neutral chemistry involving P$^+$.
PH$_3$ was first tentatively detected \citet[][]{agundez08,tez08} in the $J=1-0$ rotational transition, and then firmly \citet[][]{agundez14} detected in IRC+10216 thanks to Herschel spectra of the $J=2-1$ line.
Because PH$_3$ is expected to be a relevant P carrier owing to hydrogenation of P on dust grains (see Sect.~\ref{surface}), accurate abundance measurements of this species are extremely important to constrain chemical models.
However, radiative transfer calculations indicate a formation of PH$_3$ not necessarily confined in the inner stellar atmosphere, unlike PN, PO, HCP, and CCP, and its abundance is hard to reproduce by models \citep{agundez14}.
Finally, \citet{koelemay22} recently reported the tentative detection of SiP towards IRC+10216, whose line profile and tentative abundance are consistent with a distribution in the inner 300 stellar radii.
Its production could be due to grain destruction that releases both Si and P in the gas, but the formation/destruction pathways cannot be constrained since they are not yet included in the KIDA \citep{wakelam12} and UMIST \citep{mcelroy13} databases.

\subsection{Other environments}
\label{other}

\subsubsection{External galaxies}
\label{galaxies}

The first, and so far unique, phosphorus molecule detected in an external galaxy is PN towards the nearby starburst Galaxy NGC 253 \citep{haasler22}, in the framework of the ALMA Comprehensive High-resolution Extragalactic Molecular Inventory (ALCHEMI) project \citep{martin21}.
The PN emission arises from two giant molecular clouds in the galaxy, and it is enhanced towards the emission peak of the dust thermal continuum emission \citep{haasler22}.
Simultaneous SiO observations confirm that also in these extragalactic clouds the abundances of PN and SiO are correlated, poiting to a shocked origin of PN as in Milky Way clouds \citep[e.g.][]{lefloch16,fontani19,rivilla18,rivilla20,bergner22}.
However, while in Galactic clouds PO is found to be always more abundant than PN, upper limits on PO abundances measured towards NGC 253 suggest that the PN abundance is comparable to or larger than that of PO.
Because $\zeta$ in the central molecular zone of NGC 253 can be above $\sim 10^{-14}$~s$^{-1}$, namely at least three orders of magnitude higher than the standard interstellar value of $1.3\times 10^{-17}$~s$^{-1}$ \citep[e.g.][]{padovani09}, this difference could be due to the photo-destruction caused by secondary UV photons produced by such high $\zeta$, more efficient in destroying PO than PN \citep{jimenez18}.
The same scenario is consistent with the non-detection of PO in multiple molecular clouds of the Galactic Centre \citep{rivilla18}, also known to be associated with $\zeta$ higher by several orders of magnitude than the standard interstellar value (see Sect.~\ref{starforming}).

\subsubsection{Solar system objects and implications for astrobiology}
\label{solar}

In the Solar system, phosphorus in the form of PH$_3$ has been observed in the atmospheres of Jupiter and Saturn \citep[e.g.][respectively]{larson77,fletcher09}. 
Using an Infrared spectrometer on board of the Cassini satellite, the vertical and meridional variation of the PH$_3$ abundance in Saturn's upper troposphere was determined \citep{fletcher07}.
In the atmosphere of exoplanets, and of terrestrial planets in particular, PH$_3$ is claimed to be a promising biosignature gas \citep{sousa20} (even though it was found also in Jupiter and Saturn without necessarily being related to biological activity),
as well as an interstellar source of key phosphorus oxoacids \citep{turner18} such as phosphoric acid (H$_3$PO$_4$), phosphonic acid (H$_3$PO$_3$), and pyrophosphoric acid (H$_4$P$_2$O$_7$). 
A detection was claimed in the atmosphere of Venus \citep{greaves21} based on observations of the James Clerk Maxwell Telescope (JCMT) and ALMA, questioned by an independent analysis of the same data \citep{villanueva21}.
In meteorites, P was identified in the form of the mineral schreibersite \citep{pel05} and phosphoric acids \citep{schwartz06}.
When entering the upper atmosphere of planets, the P contained in meteoric material can be delivered at the planet surface.
At the Earth surface, the dominant form(s) of P are apatites, minerals containing P as (PO$_4)_3$, which are solid-state precipitated forms of phosphates, with Ca$^{2+}$ (or Mg$^{2+}$) and other counterions (F$^-$, Cl$^-$, OH$^-$).
However, apatites are not biologically available because poorly soluble and reactive.
Therefore, P in forms more appropriate for biological processes could have originated from such meteoric material, and been deposited at Earth surface during the early heavy bombardment period \citep{plane21}.
In comets, the measurements of the Rosetta spacecraft detected P in the comet 67P/Churyumov–Gerasimenko \citep{altwegg16}, which is predominantly in the form of phosphorus monoxide, PO \citep{rivilla20}.
Since comets, like meteorites, may have contributed to deliver at least part of the pre-biotic material to the early Earth, this finding is in line with the possibility that the basic bond of phosphates in living organisms can have a cometary origin.

\section{Input from laboratory experiments and computational chemistry}
\label{laboratory}

The synergy between observations, chemical models, laboratory experiments, and computational chemistry made it possible the identification of the aforementioned phosphorus molecules in the ISM.
Laboratory experiments, in particular, were essential to identify the new species in the observed spectra through calculation of their spectroscopic parameters.
As stated in Sect.~\ref{detection}, almost the totality of the gas-phase species in the ISM was identified through their rotational transitions at centimeter and (sub-)millimeter wavelengths, owing to the low temperatures of the ISM (see Sect.~\ref{detection}).
Therefore, rotational spectroscopy was, and still is, of paramount importance also for the identification of phosphorus molecules.
Even though this review article focusses on observations, we briefly summarise some spectroscopy works that allowed to identify the most relevant P-bearing species in centimeter and (sub-)millimeter spectra.

Several rotational transitions in the ground and first four vibrationally excited state of PN, the molecule detected so far in the highest number of sources, were first measured by \citet{wyse72} through a high temperature, microwave spectrometer.
In the same year, \citet{hoeft72} resolved the hyperfine structure of the $J=1-0$ transition in the ground and first vibrationally excited state, due to the $^{14}$N quadrupole coupling constant.
The first infrared spectrum of PN (near 1300 cm$^{-1}$) was then measured by \citet{mel81} with a tunable diode laser at high temperature, and extended to the full spectrum by \citet{aeh95}, who combined their results with previous high-temperature microwave and infrared data to produce accurate spectroscopic parameters for the ground state $X^1\Sigma^+$.
Triggered by the first PN detections in space, improved ground state rotational parameters were measured by \citet{cazzoli06}, who also derived the equilibrium structure of the molecule with high accuracy combining their measurements with rotational and ro-vibrational transitions available in the literature.
For PO, the second molecule easier to detect in the ISM, the first spectroscopic works in the far-infrared and millimeter are those of \citet{kawaguchi83} and \citet{bailleux02}, who determined also the equilibrium P-O bond length.
For the other P-bearing species detected, HCP is particularly relevant because supposed to be the main P reservoir in the inner circumstellar envelopes of C-rich stars (see Sect.~\ref{models:evolved}).
The ground state rotational spectra of HCP and its less abundant isotopologues have been measured in the millimeter-wave and submillimeter-wave ranges by \citet{johns71}, \citet{drean96}, and \citet{bizzocchi01}.
The spectroscopy of PH$_3$, believed to be the main P carrier on the surfaces of dust grains (see Sect.~\ref{surface}), was investigated by several authors \citep[e.g.][]{davies71,cep06}.
The first rotational transitions and rotational constants were estimated by \citet{heg69}.
\citet{cep06} resolved for the first time the hyperfine structure of the $J=1-0$ and $J=2-1 (K=0,1)$ rotational transitions using the Lamb-dip technique.
Some spectroscopic parameters were re-analised and improved in following works \citep{sousa13,muller13}.
Finally, the CP and CCP radicals were identified thanks to the laboratory works in \citet{saito89} and \citet{halfen08}, respectively, and the tentative detection of SiP was based on the laboratory measurements described in \citet{koelemay22}.

Finally, computational simulations were relevant to understand, from a theoretical point of view, some key aspects of chemical reactions involving P (like binding energies, reaction energies, or activation barriers) particularly challenging to study in the laboratory due to the high reactivity of phosphorus radicals.
For example, \citet{mek21} confirmed that P can be easily hydrogenated via subsequent additions of H on cold dust grain analogues, such as ice mantles.
However, using a novel methodology based on a neural network interatomic potential, \citet{molpeceres23} studied the surface reaction P + H $\rightarrow$ PH occurring on amorphous solid water, and found that the nascent PH molecule migrates rapidly towards sites with high binding energies, implying an inefficient chemical desorption, and hence very low gas-phase abundances of PH$_x$ molecules.
About the still debated topic of the prebiotic source of P on Earth (see Sect.~\ref{solar}), laboratory experiments suggested that the schreibersite mineral that reached the primordial Earth through the early meteoritic bombardment could have been converted to oxygenated phosphorus molecules upon reaction with water \citep{pel05}.
Computational studies confirmed that water corrosion of different crystalline surfaces of schreibersite indeed lead to the formation of phosphites and phosphates \citep{pantaleone21,pantaleone23}.

\section{Implications for astrochemical models: what are the main phosphorus reservoirs?}
\label{models}

Although the observational constraints obtained so far are important "food" for chemical models, the reactions leading to the formation of the detected phosphorus species are far from clear.
Moreover, a major question still remains unanswered: what are the main phosphorus reservoirs in the molecular ISM?
We will briefly summarise in this chapter the most relevant steps performed by chemical modelling to answer this question in star-forming regions (\ref{models:starforming}) and evolved stars (\ref{models:evolved}), triggered by the observational results presented in Sect.~\ref{sec:molecules}.

\subsection{Star-forming regions}
\label{models:starforming}

The fact that in all star-forming regions studied so far both PN and PO emission arise from shocked material and is well correlated with the emission of typical shock tracers (\ref{sec:molecules}), indicates very clearly dust grains as the main source of P.
However, which is the main carrier in dust grains is still debated.
\citet{jimenez18} proposed PH$_3$ on ice mantles, owing to an efficient hydrogenation of P, which however in hot or shocked gas is rapidly converted into PN and PO upon evaporation \citep[at $T\geq 100$~K,][]{chantzos20} from the grains.
But this is at odds with high-angular resolution maps  \citep{rivilla20,bergner22,fontani24}, which do not detect PN and PO towards warm/hot protostellar envelopes, where PH$_3$ should have been abundantly evaporated.
These maps thus suggest a P carrier in the refractory core of grain mantles, released by grain sputtering due to a shock passage.
PO can be efficiently formed from either PN + O \citep{jimenez18} or P + OH \cite{garcia21} if the environment is warm and/or energetic (in particular in shocks with shock speed $\geq 40$~\kms).
However, \citet{rivilla20} and \citet{bergner22} found that PN and PO emission is cospatial with low-velocity and not with high-velocity shock emission, and \citet{garcia24} found that in such low(er) velocity shocks the P + OH channel is not efficient anymore, owing to an underproduction of OH.
Rather, the most efficient formation pathway for PO production in this case should be P + O$_2$ \citep{garcia24}.

The analysis of the spatial distribution of the PN and PO emission in B1-a carried out by \citet{bergner22} suggested that there is a phosphorus reservoir in dust grains more volatile than silicate grains but less volatile than simple ice mantles (such as, for example, \METH\ ices).
Some phosphate minerals (e.g. apatite) and/or phosphorus oxides (e.g.~P$_4$O$_{10}$) possess this property, and could also explain the high abundance of PO, since they both have P-O bonds and hence could provide PO directly into the gas upon decomposition in shocks.
Such conclusion is in line with the findings of \citet{rivilla20} and \citet{bergner22}, who showed a tight spatial association between P-bearing and S-bearing molecules, suggesting S and P parent carriers with similar volatility on grain mantles.
\citet{bergner22} proposed both FeS and allotropic sulfur as plausible grain carriers of sulfur with intermediate volatilities between ices and
silicates.
However, more recently \citet{fontani24} found a spatial correspondence between PN and SiO much better than that between PN and SO or SO$_2$, which points to a P carrier more refractory than S carriers in grains.
Therefore, it is now clear that the main reservoir of P in star-forming regions is in dust grains, and cannot be in volatile ice mantles.
Nevertheless, whether it is more or less refractory with respect to some S and Si reservoirs is still under debate.

\begin{figure}[h]
    \centering
\includegraphics[width=0.8\textwidth]{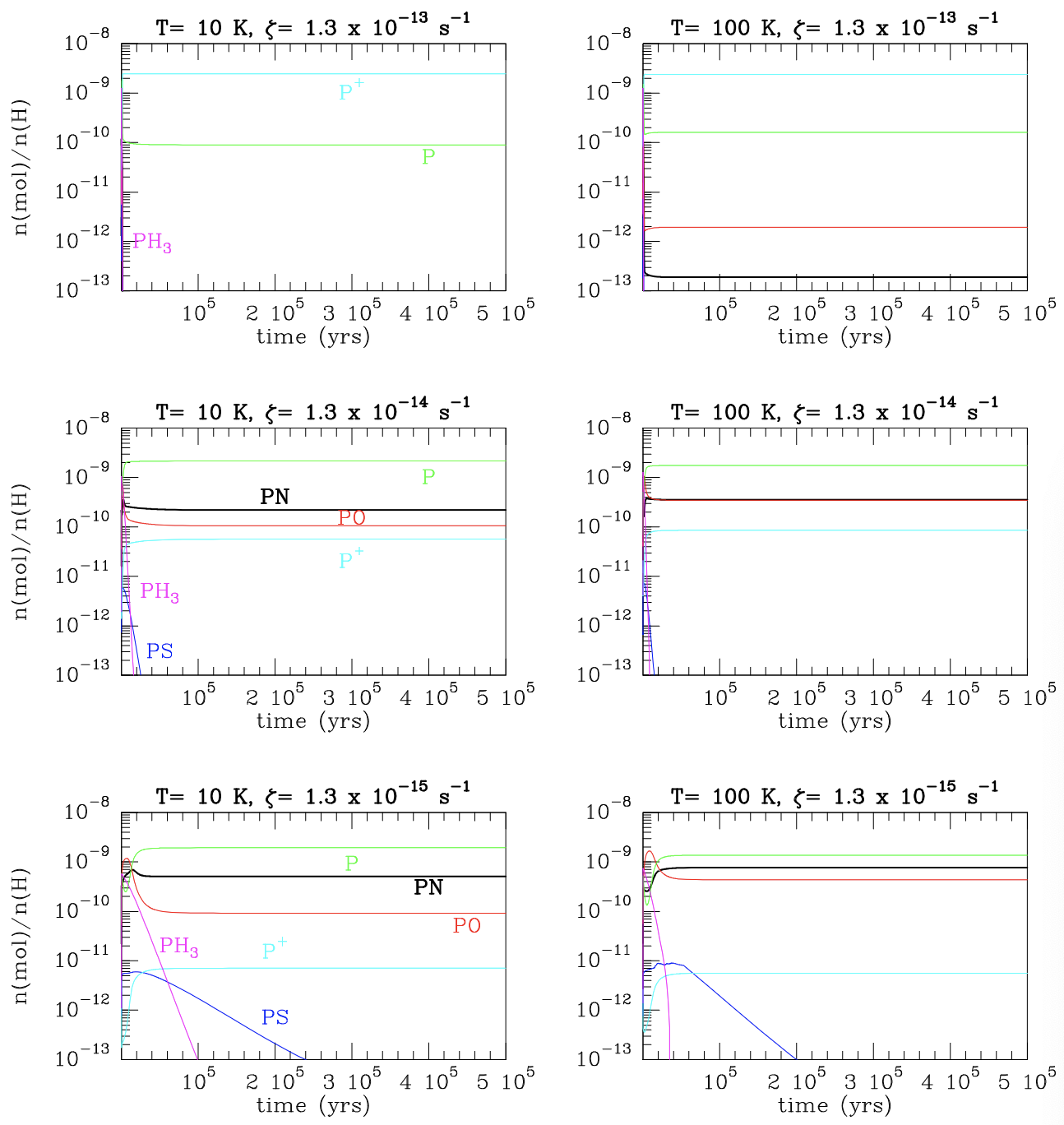}
    \caption{Evolution of the abundances of P-bearing species as a function of time in a collapsing cloud simulated for a hydrogen density n(H)=$2\times 10^4$~\cmc; 
    enhanced cosmic-ray ionisation rates $\zeta=1.3\times 10^{-15}$, $1.3\times 10^{-14}$, and $\zeta=1.3\times 10^{-13}$~s$^{-1}$; gas temperatures $T=10$ and 100~K, considering a long-lived phase for the collapse. 
    From \citet{jimenez18}.}
    \label{fig:jimenez}
\end{figure}

\citet{jimenez18} investigated how different energetic conditions can influence the P chemistry.
They concluded that models with strong UV illumination predict a strong molecular dissociation, in line with the non-detection of PO in giant molecular clouds in NGC 253 \citep{haasler22} and in the Galactic Centre \citep{rivilla18}.
Models with enhanced $\zeta$ predict high abundances of both PN and PO, but the PO/PN ratio tends to be $\leq 1$, at odds with the measured abundances.
Figure~\ref{fig:jimenez}, taken from \citet{jimenez18}, shows how the abundances of PN and PO change with $\zeta$: up to $\zeta\sim 10^{-14}$~s$^{-1}$, that is $10^{3}$ times higher than the average interstellar value, the PO/PN ratio is predicted to be $\leq 1$, while it is higher at higher $\zeta$ for a gas temperature of 100~K.
From the point of view of the chemical reactions most relevant for P chemistry, theoretical calculations of the interconversion mechanisms of PN and PO, namely N + PO and O + PN, indicate that the N + PO reaction is barrierless, while the O + PN reaction possesses a small barrier \citep{souza21}, which could hence make the difference in the PO/PN ratio in cold and warm environments.
The theoretical study of \citet{sil21} attributed the reason for the PO/PN ratio $\geq 1$ in dense and evolved cores to a lower abundance of atomic N in such environments.
\citet{fernandez23} have determined, through a mathematical model and Bayesian methods, a relatively small sample of reactions crucial in triggering P chemistry in a molecular cloud, and hence whose rates need to be constrained in the best possible way.
Depending on the temperature, the key reactions are obviously different.
At low temperature ($T\sim 10$~K), a handful of reactions should dominate the formation of PN and PO, such as \citep{fernandez23}:
\begin{equation}
\begin{split}
    {\rm O + PH} & {\rm \rightarrow PO + H} \;, \\
    {\rm O + PH_2} & {\rm \rightarrow PO + H_2} \;, \\
    {\rm N + PH} & {\rm \rightarrow PN + H} \;.
\end{split}
    \label{eq:fernandez}
\end{equation}
Because the three reactions \ref{eq:fernandez} have similar rate coefficients, in cold gas a higher PO/PN abundance would just be due to the higher O over N elemental abundance.
At higher temperatures the network becomes more complicated and more reactions, including destruction of PH and PH$_2$, need to be considered.
Moreover, \citet{fernandez23} also suggested that current astrochemical models fail in reproducing the observed PO/PN abundance ratios because some rate coefficients are likely inaccurate, and/or some networks lack important destruction routes of PN.

\subsection{Evolved stars}
\label{models:evolved}

As seen in Sect.~\ref{evolved}, phosphorus molecules are common in both oxygen- and carbon-rich evolved stars.
\citet{agundez07} modelled the envelope of the carbon-rich star IRC+20126 with thermodynamic calculations in TE conditions, and found that all phosphorus should be in the form of HCP (see Fig.~\ref{fig:agundez}).
A similar prediction was obtained by \citet{turner90}, before the first detection of HCP.
HCP should arise from a inner shell (up to 2-3 radii) around the stellar photosphere, as confirmed by observations \citet{agundez07}.
On the other hand, in an O-rich envelope the main phosphorus carriers would be PO, PS, and P$_2$ in the vicinity
of the star, and P$_4$O$_6$ at larger radii (see Fig.~\ref{fig:agundez}).
PN would be, in this case, more than 2 orders of magnitude less abundant than PO, which is clearly at odds with observational results.
To explain the observed PN abundances, this species should be efficiently formed in the outer envelope through neutral-neutral reactions like N + CP, and/or CN + P.

Most of the models developed so far assume a steady, spherically-symmetric, outflow with constant expansion velocity and mass loss rate.
\citet{mec01} modelled both C- and O-rich circumstellar envelopes in this way, assuming that the dominant forms of P at the fiducial inner radius are HCP and PS for C-rich and O-rich, respectively.
They predicted that the only species with abundance greater than $10^{-10}$ should be CP in C-rich envelopes and PO in O-rich ones.
In particular, for the O-rich case, PO should be formed by P + OH and/or PS + OH, which are most efficient at an "intermediate" distance from the stellar photosphere of $\sim 2\times 10^{16}$~cm.
\citet{ziurys18} measured in five O-rich envelopes an emission peak of PO at a distance from the star comparable to the one predicted by \citet{mec01}, but the measured abundances are 100--1000 times greater than the predicted ones.
\citet{ziurys18} concluded that the observed abundances would indicate that shocks could be important for PN formation, but PO could instead be formed by photochemical processes in the outer part of the envelope.
\citet{wem97} modelled the envelope of O-rich stars assuming that PH$_3$ is the phosphorus carrier at the fiducial inner radius. 
They found that the only phosphorus-bearing species with fractional abundances greater than $10^{-10}$ are P, PH, PH$_2$, PH$_3$ and PO.
All hydrogenated species arise for photodissociation of PH$_3$, while PO would be formed by neutral-neutral reaction of atomic O with either PH or PH$_2$.
In any case, the observed abundances cannot be reproduced.

\section{Summary, conclusions and future prospects}
\label{future}

Understanding the formation of phosphorus molecules is of great importance in astrochemistry, because P is a crucial element for the development of life as we know it.
Phosphorus is mostly created by neutron capture occurring within high-mass stars, and ejected during supernova explosions, even though recent works have invoked other sources of phosphorus to explain the detection of P-bearing molecules in the outskirt of the Galactic disk. 
In diffuse clouds, all works performed so far agree that phosphorus is essentially non depleted.
Its depletion increases with increasing gas density, although model predictions indicate depletion levels lower than previously thought to reproduce the observed molecular abundances (including upper limits).
The high uncertainty in the depletion level, and hence in the initial P elemental gaseous abundance, is certainly still one of the most critical input parameters for chemical models of dense clouds.

The first detections of PN in star-forming regions, obtained many years ago, already suggested that this species is linked to shocks.
A significant step forward in the relation between phosphorus chemistry and shocks in star-forming clouds was made in the last decade, thanks to:
\begin{itemize}
\item a significant statistical increase of PN detections obtained with single-dish telescopes towards high-mass star-forming regions, which confirmed that PN is correlated with SiO both in the abundances and line shapes in a variety of Galactic environments;
\item the first PO detections, which indicated that also this radical is likely a product of shocked gas, and is typically more abundant than PN;
\item the first surveys of PN and PO in low-mass star-forming regions, which indicated that even in the low-mass regime PO and PN are linked to shocks. Moreover, their relative ratios are in good agreement with those found in the high-mass regime, indicating similar formation/destruction pathways in low- and high-mass star-formation environments;
\item the first high-angular resolution maps of phosphorus molecules, which confirmed that the PN and PO emissions spatially overlap with those of shock tracers;
\end{itemize}
All these findings point to a solid main carrier of phosphorus in star-forming regions.
However, it is still unclear if this is mainly in a volatile or refractory form.
Theory indicates that the main volatile form would be PH$_3$ in ice mantles of dust grains, but PH$_3$ was never detected in a star-forming regions so far, and the expected products of its destruction in hot gas, namely PN and PO, are also undetected towards warm/hot protostellar envelopes.
Therefore, PH$_3$ does not seem to be the main solid phosphorus carrier.
Moreover, the PN and PO emission seems to arise from material at lower velocity, on average, that that responsible for the SiO and SO emission, which
challenges the simple grain sputtering scenario.
A solid carrier of P not as volatile as ice mantles, and not as refractory as silicate cores, could solve this problem. 
However, the non-detection of PN and/or PO at high velocities could also be due to the limited sensitivity of the observations in such velocity ranges.
In any case, PO seems the main gaseous P carrier in star-forming regions.
It is also the main carrier in the comet 67P/Churyumov-Gerasimenko, which suggests that comets could have contributed significantly to the phosphorus budget (in particular in the form of phosphates) at the early evolution of Earth.

From the side of evolved stars, CP was the first molecule detected in the shell of a C-rich AGB star, followed by PN, HCP, CCP, PO, and PH$_3$.
Some of these species, in particular several lines of PN and PO, were detected also in O-rich envelopes of super-giant stars and AGB stars.
A recent tentative detection of SiP was also reported in the circumstellar shell of the C-rich envelope of IRC+10216.
These studies indicate that phosphorus molecules are common in both C-rich and O-rich envelopes of evolved stars.
Models suggest that the molecular emission arises predominantly from inner shells, and hence such molecules are created close to the stellar photosphere, where the gas is in TE conditions.
The formation of PO is apparently favoured with respect to that of PN in O-rich envelopes, where theory predicts that PO is indeed one of the main carriers.
Instead, models of C-rich stellar envelopes successfully predict that P would be mostly in the form of HCP, as observed, but the first confirmed detection of PH$_3$ in IRC+10216 challenged these models, since no obvious formation routes in TE are able to match its observed abundance.
Several non-equilibrium processes, again invoking shocks, were proposed, but in any case the observed abundances suggest gas phase carriers of P more abundant than previously thought. 

The detection of phosphorus compounds in new sources and new molecules, especially those predicted to be relevant from models such as PS in the envelope of O-rich stars, HCP and CP in diffuse clouds, and PS, HPO, and HPO$^+$ in dense star-forming cores,
will significantly help us to constrain much better the phosphorus chemistry in the ISM.
These detections will require high sensitivity observations.
The next generation of radiotelescopes, in particular of the Square Kilometer Array (SKA), will certainly play an important role.
The sensitivity of SKA in the centimeter range will allow us to detect the low-$J$ transitions of phosphorus molecules with an unprecedented signal-to-noise ratio.
Moreover, new high-angular resolution studies, both with current and future interferometers, will be critical to understand the physical mechanisms that are responsible of the formation (and destruction) of P-bearing species in the ISM. 
Obviously, such observations will benefit from further development of theoretical studies and laboratory experiments, and vice versa. 
New chemical routes should be explored, and their associated reaction rates should be measured with great accuracy.
Laboratory experiments should derive accurate spectroscopic parameters for proper identification of molecular lines.
In summary, the synergy between new observations, model predictions, laboratory measurements, and computational simulations will allow us to further progress in our knowledge of the important but still mysterious interstellar chemistry of phosphorus.

\tiny
 \keyFont{ \section{Keywords:} star formation, interstellar medium, astrochemistry, protostars, evolved stars, } %All article types: you may provide up to 8 keywords; at least 5 are mandatory.

\section*{Conflict of Interest Statement}
%All financial, commercial or other relationships that might be perceived by the academic community as representing a potential conflict of interest must be disclosed. If no such relationship exists, authors will be asked to confirm the following statement: 

The author declares that the research was conducted in the absence of any commercial or financial relationships that could be construed as a potential conflict of interest.

\section*{Author Contributions}

FF has totally written this manuscript

%\section*{Funding}
%Details of all funding sources should be provided, including grant numbers if applicable. Please ensure to add all necessary funding information, as after publication this is no longer possible.

\section*{Acknowledgments}

F.F. is grateful to the reviewers for their constructive comments, which helped to improve the original version of the manuscript.
F.F. is also grateful to V. Lattanzi, V.M. Rivilla, and L. Magrini, for useful discussion.

%\section*{Supplemental Data}
% \href{http://home.frontiersin.org/about/author-guidelines#SupplementaryMaterial}{Supplementary Material} should be uploaded separately on submission, if there are Supplementary Figures, please include the caption in the same file as the figure. LaTeX Supplementary Material templates can be found in the Frontiers LaTeX folder.

%\section*{Data Availability Statement}
%The datasets [GENERATED/ANALYZED] for this study can be found in the [NAME OF REPOSITORY] [LINK].
% Please see the availability of data guidelines for more information, at https://www.frontiersin.org/about/author-guidelines#AvailabilityofData

\bibliographystyle{Frontiers-Harvard}
%  Many Frontiers journals use the Harvard referencing system (Author-date), to find the style and resources for the journal you are submitting to: https://zendesk.frontiersin.org/hc/en-us/articles/360017860337-Frontiers-Reference-Styles-by-Journal. For Humanities and Social Sciences articles please include page numbers in the in-text citations 
%\bibliographystyle{Frontiers-Vancouver} % Many Frontiers journals use the numbered referencing system, to find the style and resources for the journal you are submitting to: https://zendesk.frontiersin.org/hc/en-us/articles/360017860337-Frontiers-Reference-Styles-by-Journal
\bibliography{Phosphorus-review}

\begin{thebibliography}{140}
\providecommand{\natexlab}[1]{#1}
\expandafter\ifx\csname urlstyle\endcsname\relax
  \providecommand{\doi}[1]{doi:\discretionary{}{}{}#1}\else
  \providecommand{\doi}{doi:\discretionary{}{}{}\begingroup
  \urlstyle{rm}\Url}\fi
\providecommand{\selectlanguage}[1]{\relax}
\providecommand{\bibAnnoteFile}[1]{%
  \IfFileExists{#1}{\begin{quotation}\noindent\textsc{Key:} #1\\
  \textsc{Annotation:}\ \input{#1}\end{quotation}}{}}
\providecommand{\bibAnnote}[2]{%
  \begin{quotation}\noindent\textsc{Key:} #1\\
  \textsc{Annotation:}\ #2\end{quotation}}

\bibitem[{{Adams} et~al.(1990){Adams}, {McIntosh}, and {Smith}}]{adams90}
{Adams}, N.~G., {McIntosh}, B.~J., and {Smith}, D. (1990).
\newblock {Production of phosphorus-containing molecules in interstellar
  clouds.}
\newblock \emph{\aap} 232, 443
\bibAnnoteFile{adams90}

\bibitem[{{Ag{\'u}ndez} et~al.(2014){Ag{\'u}ndez}, {Cernicharo}, {Decin},
  {Encrenaz}, and {Teyssier}}]{agundez14}
{Ag{\'u}ndez}, M., {Cernicharo}, J., {Decin}, L., {Encrenaz}, P., and
  {Teyssier}, D. (2014).
\newblock {Confirmation of Circumstellar Phosphine}.
\newblock \emph{\apjl} 790, L27.
\newblock \doi{10.1088/2041-8205/790/2/L27}
\bibAnnoteFile{agundez14}

\bibitem[{{Ag{\'u}ndez} et~al.(2007){Ag{\'u}ndez}, {Cernicharo}, and
  {Gu{\'e}lin}}]{agundez07}
{Ag{\'u}ndez}, M., {Cernicharo}, J., and {Gu{\'e}lin}, M. (2007).
\newblock {Discovery of Phosphaethyne (HCP) in Space: Phosphorus Chemistry in
  Circumstellar Envelopes}.
\newblock \emph{\apjl} 662, L91--L94.
\newblock \doi{10.1086/519561}
\bibAnnoteFile{agundez07}

\bibitem[{{Ag{\'u}ndez} et~al.(2008){Ag{\'u}ndez}, {Cernicharo}, {Pardo},
  {Gu{\'e}lin}, and {Phillips}}]{agundez08}
{Ag{\'u}ndez}, M., {Cernicharo}, J., {Pardo}, J.~R., {Gu{\'e}lin}, M., and
  {Phillips}, T.~G. (2008).
\newblock {Tentative detection of phosphine in IRC +10216}.
\newblock \emph{\aap} 485, L33--L36.
\newblock \doi{10.1051/0004-6361:200810193}
\bibAnnoteFile{agundez08}

\bibitem[{{Ahmad} and {Hamilton}(1995)}]{aeh95}
{Ahmad}, I.~K. and {Hamilton}, P.~A. (1995).
\newblock {The Fourier Transform Infrared Spectrum of PN}.
\newblock \emph{Journal of Molecular Spectroscopy} 169, 286--291.
\newblock \doi{10.1006/jmsp.1995.1022}
\bibAnnoteFile{aeh95}

\bibitem[{{Altwegg} et~al.(2016){Altwegg}, {Balsiger}, {Bar-Nun}, {Berthelier},
  {Bieler}, {Bochsler} et~al.}]{altwegg16}
{Altwegg}, K., {Balsiger}, H., {Bar-Nun}, A., {Berthelier}, J.~J., {Bieler},
  A., {Bochsler}, P., et~al. (2016).
\newblock {Prebiotic chemicals--amino acid and phosphorus--in the coma of comet
  67P/Churyumov-Gerasimenko}.
\newblock \emph{Science Advances} 2, e1600285--e1600285.
\newblock \doi{10.1126/sciadv.1600285}
\bibAnnoteFile{altwegg16}

\bibitem[{{Anders} and {Grevesse}(1989)}]{aeg89}
{Anders}, E. and {Grevesse}, N. (1989).
\newblock {Abundances of the elements: Meteoritic and solar}.
\newblock \emph{\gca} 53, 197--214.
\newblock \doi{10.1016/0016-7037(89)90286-X}
\bibAnnoteFile{aeg89}

\bibitem[{{Anicich}(1993)}]{anicich93}
{Anicich}, V.~G. (1993).
\newblock {A Survey of Bimolecular Ion-Molecule Reactions for Use in Modeling
  the Chemistry of Planetary Atmospheres, Cometary Comae, and Interstellar
  Clouds: 1993 Supplement}.
\newblock \emph{\apjs} 84, 215.
\newblock \doi{10.1086/191752}
\bibAnnoteFile{anicich93}

\bibitem[{{Asplund} et~al.(2009){Asplund}, {Grevesse}, {Sauval}, and
  {Scott}}]{asplund09}
{Asplund}, M., {Grevesse}, N., {Sauval}, A.~J., and {Scott}, P. (2009).
\newblock {The Chemical Composition of the Sun}.
\newblock \emph{\araa} 47, 481--522.
\newblock \doi{10.1146/annurev.astro.46.060407.145222}
\bibAnnoteFile{asplund09}

\bibitem[{{Bailleux} et~al.(2002){Bailleux}, {Bogey}, {Demuynck}, {Liu}, and
  {Walters}}]{bailleux02}
{Bailleux}, S., {Bogey}, M., {Demuynck}, C., {Liu}, Y., and {Walters}, A.
  (2002).
\newblock {Millimeter-Wave Spectroscopy of PO in Excited Vibrational States up
  to v=7}.
\newblock \emph{Journal of Molecular Spectroscopy} 216, 465--471.
\newblock \doi{10.1006/jmsp.2002.8665}
\bibAnnoteFile{bailleux02}

\bibitem[{{Bekki} and {Tsujimoto}(2024)}]{bet24}
{Bekki}, K. and {Tsujimoto}, T. (2024).
\newblock {Phosphorus Enrichment by ONe Novae in the Galaxy}.
\newblock \emph{\apjl} 967, L1.
\newblock \doi{10.3847/2041-8213/ad3fb6}
\bibAnnoteFile{bet24}

\bibitem[{{Bergner} et~al.(2022){Bergner}, {Burkhardt}, {{\"O}berg}, {Rice},
  and {Bergin}}]{bergner22}
{Bergner}, J.~B., {Burkhardt}, A.~M., {{\"O}berg}, K.~I., {Rice}, T.~S., and
  {Bergin}, E.~A. (2022).
\newblock {First Images of Phosphorus Molecules toward a Protosolar Analog}.
\newblock \emph{\apj} 927, 7.
\newblock \doi{10.3847/1538-4357/ac47a2}
\bibAnnoteFile{bergner22}

\bibitem[{{Bergner} et~al.(2019){Bergner}, {{\"O}berg}, {Walker}, {Guzm{\'a}n},
  {Rice}, and {Bergin}}]{bergner19}
{Bergner}, J.~B., {{\"O}berg}, K.~I., {Walker}, S., {Guzm{\'a}n}, V.~V.,
  {Rice}, T.~S., and {Bergin}, E.~A. (2019).
\newblock {Detection of Phosphorus-bearing Molecules toward a Solar-type
  Protostar}.
\newblock \emph{\apjl} 884, L36.
\newblock \doi{10.3847/2041-8213/ab48f9}
\bibAnnoteFile{bergner19}

\bibitem[{{Bizzocchi} et~al.(2001){Bizzocchi}, {Thorwirth}, {M{\"u}ller},
  {Lewen}, and {Winnewisser}}]{bizzocchi01}
{Bizzocchi}, L., {Thorwirth}, S., {M{\"u}ller}, H. S.~P., {Lewen}, F., and
  {Winnewisser}, G. (2001).
\newblock {Submillimeter-Wave Spectroscopy of Phosphaalkynes: HCCCP, NCCP, HCP,
  and DCP}.
\newblock \emph{Journal of Molecular Spectroscopy} 205, 110--116.
\newblock \doi{10.1006/jmsp.2000.8234}
\bibAnnoteFile{bizzocchi01}

\bibitem[{{Blair} et~al.(2008){Blair}, {Magnani}, {Brand}, and
  {Wouterloot}}]{blair08}
{Blair}, S.~K., {Magnani}, L., {Brand}, J., and {Wouterloot}, J. G.~A. (2008).
\newblock {Formaldehyde in the Far Outer Galaxy: Constraining the Outer
  Boundary of the Galactic Habitable Zone}.
\newblock \emph{Astrobiology} 8, 59--73.
\newblock \doi{10.1089/ast.2007.0171}
\bibAnnoteFile{blair08}

\bibitem[{{Boogert} et~al.(2015){Boogert}, {Gerakines}, and
  {Whittet}}]{boogert15}
{Boogert}, A.~C.~A., {Gerakines}, P.~A., and {Whittet}, D. C.~B. (2015).
\newblock {Observations of the icy universe.}
\newblock \emph{\araa} 53, 541--581.
\newblock \doi{10.1146/annurev-astro-082214-122348}
\bibAnnoteFile{boogert15}

\bibitem[{{Caffau} et~al.(2011){Caffau}, {Bonifacio}, {Faraggiana}, and
  {Steffen}}]{caffau11}
{Caffau}, E., {Bonifacio}, P., {Faraggiana}, R., and {Steffen}, M. (2011).
\newblock {The Galactic evolution of phosphorus}.
\newblock \emph{\aap} 532, A98.
\newblock \doi{10.1051/0004-6361/201117313}
\bibAnnoteFile{caffau11}

\bibitem[{{Calmonte} et~al.(2016){Calmonte}, {Altwegg}, {Balsiger},
  {Berthelier}, {Bieler}, {Cessateur} et~al.}]{calmonte16}
{Calmonte}, U., {Altwegg}, K., {Balsiger}, H., {Berthelier}, J.~J., {Bieler},
  A., {Cessateur}, G., et~al. (2016).
\newblock {Sulphur-bearing species in the coma of comet
  67P/Churyumov-Gerasimenko}.
\newblock \emph{\mnras} 462, S253--S273.
\newblock \doi{10.1093/mnras/stw2601}
\bibAnnoteFile{calmonte16}

\bibitem[{{Caselli} and {Ceccarelli}(2012)}]{cec12}
{Caselli}, P. and {Ceccarelli}, C. (2012).
\newblock {Our astrochemical heritage}.
\newblock \emph{\aapr} 20, 56.
\newblock \doi{10.1007/s00159-012-0056-x}
\bibAnnoteFile{cec12}

\bibitem[{{Cazaux} and {Tielens}(2004)}]{cet04}
{Cazaux}, S. and {Tielens}, A.~G.~G.~M. (2004).
\newblock {H$_{2}$ Formation on Grain Surfaces}.
\newblock \emph{\apj} 604, 222--237.
\newblock \doi{10.1086/381775}
\bibAnnoteFile{cet04}

\bibitem[{{Cazzoli} et~al.(2006){Cazzoli}, {Cludi}, and
  {Puzzarini}}]{cazzoli06}
{Cazzoli}, G., {Cludi}, L., and {Puzzarini}, C. (2006).
\newblock {Microwave spectrum of P $^{14}$N and P $^{15}$N: Spectroscopic
  constants and molecular structure}.
\newblock \emph{Journal of Molecular Structure} 780, 260--267.
\newblock \doi{10.1016/j.molstruc.2005.07.010}
\bibAnnoteFile{cazzoli06}

\bibitem[{{Cazzoli} and {Puzzarini}(2006)}]{cep06}
{Cazzoli}, G. and {Puzzarini}, C. (2006).
\newblock {The Lamb-dip spectrum of phosphine: The nuclear hyperfine structure
  due to hydrogen and phosphorus}.
\newblock \emph{Journal of Molecular Spectroscopy} 239, 64--70.
\newblock \doi{10.1016/j.jms.2006.05.019}
\bibAnnoteFile{cep06}

\bibitem[{{Ceccarelli} et~al.(2023){Ceccarelli}, {Codella}, {Balucani},
  {Bockelee-Morvan}, {Herbst}, {Vastel} et~al.}]{ceccarelli23}
{Ceccarelli}, C., {Codella}, C., {Balucani}, N., {Bockelee-Morvan}, D.,
  {Herbst}, E., {Vastel}, C., et~al. (2023).
\newblock {Organic Chemistry in the First Phases of Solar-Type Protostars}.
\newblock In \emph{Protostars and Planets VII}, eds. S.~{Inutsuka},
  Y.~{Aikawa}, T.~{Muto}, K.~{Tomida}, and M.~{Tamura}. vol. 534 of
  \emph{Astronomical Society of the Pacific Conference Series}, 379
\bibAnnoteFile{ceccarelli23}

\bibitem[{{Cernicharo} et~al.(2000){Cernicharo}, {Gu{\'e}lin}, and
  {Kahane}}]{cernicharo00}
{Cernicharo}, J., {Gu{\'e}lin}, M., and {Kahane}, C. (2000).
\newblock {A lambda 2 mm molecular line survey of the C-star envelope
  IRC+10216}.
\newblock \emph{\aaps} 142, 181--215.
\newblock \doi{10.1051/aas:2000147}
\bibAnnoteFile{cernicharo00}

\bibitem[{{Chantzos} et~al.(2020){Chantzos}, {Rivilla}, {Vasyunin}, {Redaelli},
  {Bizzocchi}, {Fontani} et~al.}]{chantzos20}
{Chantzos}, J., {Rivilla}, V.~M., {Vasyunin}, A., {Redaelli}, E., {Bizzocchi},
  L., {Fontani}, F., et~al. (2020).
\newblock {The first steps of interstellar phosphorus chemistry}.
\newblock \emph{\aap} 633, A54.
\newblock \doi{10.1051/0004-6361/201936531}
\bibAnnoteFile{chantzos20}

\bibitem[{{Charnley} and {Millar}(1994)}]{cem94}
{Charnley}, S.~B. and {Millar}, T.~J. (1994).
\newblock {The Chemistry of Phosphorus in Hot Molecular Cores}.
\newblock \emph{\mnras} 270, 570.
\newblock \doi{10.1093/mnras/270.3.570}
\bibAnnoteFile{cem94}

\bibitem[{{Cherchneff}(2006)}]{cherchneff06}
{Cherchneff}, I. (2006).
\newblock {A chemical study of the inner winds of asymptotic giant branch
  stars}.
\newblock \emph{\aap} 456, 1001--1012.
\newblock \doi{10.1051/0004-6361:20064827}
\bibAnnoteFile{cherchneff06}

\bibitem[{{Clayton}(2003)}]{clayton03}
{Clayton}, D. (2003).
\newblock \emph{{Handbook of Isotopes in the Cosmos}} (Cambridge University
  Press)
\bibAnnoteFile{clayton03}

\bibitem[{{Davies} et~al.(1971){Davies}, {Neumann}, {Wofsy}, and
  {Klemperer}}]{davies71}
{Davies}, P.~B., {Neumann}, R.~M., {Wofsy}, S.~C., and {Klemperer}, W. (1971).
\newblock {Radio-Frequency Spectrum of Phosphine (PH$_{3}$)}.
\newblock \emph{\jcp} 55, 3564--3568.
\newblock \doi{10.1063/1.1676614}
\bibAnnoteFile{davies71}

\bibitem[{{De Beck} et~al.(2013){De Beck}, {Kami{\'n}ski}, {Patel}, {Young},
  {Gottlieb}, {Menten} et~al.}]{debeck13}
{De Beck}, E., {Kami{\'n}ski}, T., {Patel}, N.~A., {Young}, K.~H., {Gottlieb},
  C.~A., {Menten}, K.~M., et~al. (2013).
\newblock {PO and PN in the wind of the oxygen-rich AGB star IK Tauri}.
\newblock \emph{\aap} 558, A132.
\newblock \doi{10.1051/0004-6361/201321349}
\bibAnnoteFile{debeck13}

\bibitem[{{Draine}(2011)}]{draine11}
{Draine}, B.~T. (2011).
\newblock \emph{{Physics of the Interstellar and Intergalactic Medium}}
  (Published by Princeton University Press)
\bibAnnoteFile{draine11}

\bibitem[{{Draine} and {Li}(2001)}]{del01}
{Draine}, B.~T. and {Li}, A. (2001).
\newblock {Infrared Emission from Interstellar Dust. I. Stochastic Heating of
  Small Grains}.
\newblock \emph{\apj} 551, 807--824.
\newblock \doi{10.1086/320227}
\bibAnnoteFile{del01}

\bibitem[{{Dr{\'e}an} et~al.(1996){Dr{\'e}an}, {Demaison}, {Poteau}, and
  {Denis}}]{drean96}
{Dr{\'e}an}, P., {Demaison}, J., {Poteau}, L., and {Denis}, J.~M. (1996).
\newblock {Rotational Spectrum and Structure of HCP}.
\newblock \emph{Journal of Molecular Spectroscopy} 176, 139--145.
\newblock \doi{10.1006/jmsp.1996.0070}
\bibAnnoteFile{drean96}

\bibitem[{{Dufton} et~al.(1986){Dufton}, {Keenan}, and {Hibbert}}]{dufton86}
{Dufton}, P.~L., {Keenan}, F.~P., and {Hibbert}, A. (1986).
\newblock {The abundance of phosphorus in the interstellar medium.}
\newblock \emph{\aap} 164, 179--183
\bibAnnoteFile{dufton86}

\bibitem[{{Duley} and {Williams}(1985)}]{dew85}
{Duley}, W.~W. and {Williams}, D.~A. (1985).
\newblock {Book-Review - Interstellar Chemistry}.
\newblock \emph{Science} 227, 786
\bibAnnoteFile{dew85}

\bibitem[{{Endres} et~al.(2016){Endres}, {Schlemmer}, {Schilke}, {Stutzki}, and
  {M{\"u}ller}}]{endres16}
{Endres}, C.~P., {Schlemmer}, S., {Schilke}, P., {Stutzki}, J., and
  {M{\"u}ller}, H. S.~P. (2016).
\newblock {The Cologne Database for Molecular Spectroscopy, CDMS, in the
  Virtual Atomic and Molecular Data Centre, VAMDC}.
\newblock \emph{Journal of Molecular Spectroscopy} 327, 95--104.
\newblock \doi{10.1016/j.jms.2016.03.005}
\bibAnnoteFile{endres16}

\bibitem[{{Fern{\'a}ndez-Ruz} et~al.(2023){Fern{\'a}ndez-Ruz},
  {Jim{\'e}nez-Serra}, and {Aguirre}}]{fernandez23}
{Fern{\'a}ndez-Ruz}, M., {Jim{\'e}nez-Serra}, I., and {Aguirre}, J. (2023).
\newblock {A Theoretical Approach to the Complex Chemical Evolution of
  Phosphorus in the Interstellar Medium}.
\newblock \emph{\apj} 956, 47.
\newblock \doi{10.3847/1538-4357/acf290}
\bibAnnoteFile{fernandez23}

\bibitem[{{Fletcher} et~al.(2007){Fletcher}, {Irwin}, {Teanby}, {Orton},
  {Parrish}, {Calcutt} et~al.}]{fletcher07}
{Fletcher}, L.~N., {Irwin}, P.~G.~J., {Teanby}, N.~A., {Orton}, G.~S.,
  {Parrish}, P.~D., {Calcutt}, S.~B., et~al. (2007).
\newblock {The meridional phosphine distribution in Saturn's upper troposphere
  from Cassini/CIRS observations}.
\newblock \emph{\icarus} 188, 72--88.
\newblock \doi{10.1016/j.icarus.2006.10.029}
\bibAnnoteFile{fletcher07}

\bibitem[{{Fletcher} et~al.(2009){Fletcher}, {Orton}, {Teanby}, and
  {Irwin}}]{fletcher09}
{Fletcher}, L.~N., {Orton}, G.~S., {Teanby}, N.~A., and {Irwin}, P.~G.~J.
  (2009).
\newblock {Phosphine on Jupiter and Saturn from Cassini/CIRS}.
\newblock \emph{\icarus} 202, 543--564.
\newblock \doi{10.1016/j.icarus.2009.03.023}
\bibAnnoteFile{fletcher09}

\bibitem[{{Fontani} et~al.(2022){Fontani}, {Colzi}, {Bizzocchi}, {Rivilla},
  {Elia}, {Beltr{\'a}n} et~al.}]{fontani22}
{Fontani}, F., {Colzi}, L., {Bizzocchi}, L., {Rivilla}, V.~M., {Elia}, D.,
  {Beltr{\'a}n}, M.~T., et~al. (2022).
\newblock {CHEMOUT: CHEMical complexity in star-forming regions of the OUTer
  Galaxy. I. Organic molecules and tracers of star-formation activity}.
\newblock \emph{\aap} 660, A76.
\newblock \doi{10.1051/0004-6361/202142923}
\bibAnnoteFile{fontani22}

\bibitem[{{Fontani} et~al.(2024){Fontani}, {Mininni}, {Beltr{\'a}n}, {Rivilla},
  {Colzi}, {Jim{\'e}nez-Serra} et~al.}]{fontani24}
{Fontani}, F., {Mininni}, C., {Beltr{\'a}n}, M.~T., {Rivilla}, V.~M., {Colzi},
  L., {Jim{\'e}nez-Serra}, I., et~al. (2024).
\newblock {The GUAPOS project: G31.41+0.31 Unbiased ALMA sPectral Observational
  Survey. IV. Phosphorus-bearing molecules and their relation to shock
  tracers}.
\newblock \emph{\aap} 682, A74.
\newblock \doi{10.1051/0004-6361/202348219}
\bibAnnoteFile{fontani24}

\bibitem[{{Fontani} et~al.(2016){Fontani}, {Rivilla}, {Caselli}, {Vasyunin},
  and {Palau}}]{fontani16}
{Fontani}, F., {Rivilla}, V.~M., {Caselli}, P., {Vasyunin}, A., and {Palau}, A.
  (2016).
\newblock {Phosphorus-bearing Molecules in Massive Dense Cores}.
\newblock \emph{\apjl} 822, L30.
\newblock \doi{10.3847/2041-8205/822/2/L30}
\bibAnnoteFile{fontani16}

\bibitem[{{Fontani} et~al.(2019){Fontani}, {Rivilla}, {van der Tak}, {Mininni},
  {Beltr{\'a}n}, and {Caselli}}]{fontani19}
{Fontani}, F., {Rivilla}, V.~M., {van der Tak}, F.~F.~S., {Mininni}, C.,
  {Beltr{\'a}n}, M.~T., and {Caselli}, P. (2019).
\newblock {Origin of the PN molecule in star-forming regions: the enlarged
  sample}.
\newblock \emph{\mnras} 489, 4530--4542.
\newblock \doi{10.1093/mnras/stz2446}
\bibAnnoteFile{fontani19}

\bibitem[{{Fontani} et~al.(2023){Fontani}, {Roueff}, {Colzi}, and
  {Caselli}}]{fontani23}
{Fontani}, F., {Roueff}, E., {Colzi}, L., and {Caselli}, P. (2023).
\newblock {The evolution of sulphur-bearing molecules in high-mass star-forming
  cores}.
\newblock \emph{\aap} 680, A58.
\newblock \doi{10.1051/0004-6361/202347565}
\bibAnnoteFile{fontani23}

\bibitem[{{Friedman} et~al.(2000){Friedman}, {Howk}, {Andersson}, {Sembach},
  {Ake}, {Roth} et~al.}]{friedman00}
{Friedman}, S.~D., {Howk}, J.~C., {Andersson}, B.~G., {Sembach}, K.~R., {Ake},
  T.~B., {Roth}, K., et~al. (2000).
\newblock {Far Ultraviolet Spectroscopic Explorer Observations of Interstellar
  Gas toward the Large Magellanic Cloud Star SK -67{\textdegree}05}.
\newblock \emph{\apjl} 538, L39--L42.
\newblock \doi{10.1086/312800}
\bibAnnoteFile{friedman00}

\bibitem[{{Fuente} et~al.(2023){Fuente}, {Rivi{\`e}re-Marichalar},
  {Beitia-Antero}, {Caselli}, {Wakelam}, {Esplugues} et~al.}]{fuente23}
{Fuente}, A., {Rivi{\`e}re-Marichalar}, P., {Beitia-Antero}, L., {Caselli}, P.,
  {Wakelam}, V., {Esplugues}, G., et~al. (2023).
\newblock {Gas phase Elemental abundances in Molecular cloudS (GEMS). VII.
  Sulfur elemental abundance}.
\newblock \emph{\aap} 670, A114.
\newblock \doi{10.1051/0004-6361/202244843}
\bibAnnoteFile{fuente23}

\bibitem[{{Furuya} et~al.(2022){Furuya}, {Oba}, and {Shimonishi}}]{furuya22}
{Furuya}, K., {Oba}, Y., and {Shimonishi}, T. (2022).
\newblock {Quantifying the Chemical Desorption of H$_{2}$S and PH$_{3}$ from
  Amorphous Water-ice Surfaces}.
\newblock \emph{\apj} 926, 171.
\newblock \doi{10.3847/1538-4357/ac4260}
\bibAnnoteFile{furuya22}

\bibitem[{{Furuya} and {Shimonishi}(2024)}]{fes24}
{Furuya}, K. and {Shimonishi}, T. (2024).
\newblock {Deep Search for Phosphine in a Prestellar Core}.
\newblock \emph{arXiv e-prints} ,
  arXiv:2406.05978\doi{10.48550/arXiv.2406.05978}
\bibAnnoteFile{fes24}

\bibitem[{{Garc{\'\i}a de la Concepci{\'o}n} et~al.(2024){Garc{\'\i}a de la
  Concepci{\'o}n}, {Cavallotti}, {Barone}, {Puzzarini}, and
  {Jim{\'e}nez-Serra}}]{garcia24}
{Garc{\'\i}a de la Concepci{\'o}n}, J., {Cavallotti}, C., {Barone}, V.,
  {Puzzarini}, C., and {Jim{\'e}nez-Serra}, I. (2024).
\newblock {Relevance of the P+O$_{2}$ Reaction for PO Formation in
  Astrochemical Environments: Electronic Structure Calculations and Kinetic
  Simulations}.
\newblock \emph{\apj} 963, 142.
\newblock \doi{10.3847/1538-4357/ad1ffa}
\bibAnnoteFile{garcia24}

\bibitem[{{Garc{\'\i}a de la Concepci{\'o}n} et~al.(2021){Garc{\'\i}a de la
  Concepci{\'o}n}, {Puzzarini}, {Barone}, {Jim{\'e}nez-Serra}, and
  {Roncero}}]{garcia21}
{Garc{\'\i}a de la Concepci{\'o}n}, J., {Puzzarini}, C., {Barone}, V.,
  {Jim{\'e}nez-Serra}, I., and {Roncero}, O. (2021).
\newblock {Formation of Phosphorus Monoxide (PO) in the Interstellar Medium:
  Insights from Quantum-chemical and Kinetic Calculations}.
\newblock \emph{\apj} 922, 169.
\newblock \doi{10.3847/1538-4357/ac1e94}
\bibAnnoteFile{garcia21}

\bibitem[{{Glassgold} et~al.(1987){Glassgold}, {Mamon}, {Omont}, and
  {Lucas}}]{glassgold87}
{Glassgold}, A.~E., {Mamon}, G.~A., {Omont}, A., and {Lucas}, R. (1987).
\newblock {Photochemistry and molecular ions in carbon-rich circumstellar
  envelopes.}
\newblock \emph{\aap} 180, 183--190
\bibAnnoteFile{glassgold87}

\bibitem[{{Gould} and {Salpeter}(1963)}]{ges63}
{Gould}, R.~J. and {Salpeter}, E.~E. (1963).
\newblock {The Interstellar Abundance of the Hydrogen Molecule. I. Basic
  Processes.}
\newblock \emph{\apj} 138, 393.
\newblock \doi{10.1086/147654}
\bibAnnoteFile{ges63}

\bibitem[{{Greaves} et~al.(2021){Greaves}, {Richards}, {Bains}, {Rimmer},
  {Sagawa}, {Clements} et~al.}]{greaves21}
{Greaves}, J.~S., {Richards}, A. M.~S., {Bains}, W., {Rimmer}, P.~B., {Sagawa},
  H., {Clements}, D.~L., et~al. (2021).
\newblock {Phosphine gas in the cloud decks of Venus}.
\newblock \emph{Nature Astronomy} 5, 655--664.
\newblock \doi{10.1038/s41550-020-1174-4}
\bibAnnoteFile{greaves21}

\bibitem[{{Guelin} et~al.(1990){Guelin}, {Cernicharo}, {Paubert}, and
  {Turner}}]{guelin90}
{Guelin}, M., {Cernicharo}, J., {Paubert}, G., and {Turner}, B.~E. (1990).
\newblock {Free CP in IRC +10216.}
\newblock \emph{\aap} 230, L9--L11
\bibAnnoteFile{guelin90}

\bibitem[{{Haasler} et~al.(2022){Haasler}, {Rivilla}, {Mart{\'\i}n},
  {Holdship}, {Viti}, {Harada} et~al.}]{haasler22}
{Haasler}, D., {Rivilla}, V.~M., {Mart{\'\i}n}, S., {Holdship}, J., {Viti}, S.,
  {Harada}, N., et~al. (2022).
\newblock {First extragalactic detection of a phosphorus-bearing molecule with
  ALCHEMI: Phosphorus nitride (PN)}.
\newblock \emph{\aap} 659, A158.
\newblock \doi{10.1051/0004-6361/202142032}
\bibAnnoteFile{haasler22}

\bibitem[{{Halfen} et~al.(2008){Halfen}, {Clouthier}, and {Ziurys}}]{halfen08}
{Halfen}, D.~T., {Clouthier}, D.~J., and {Ziurys}, L.~M. (2008).
\newblock {Detection of the CCP Radical (X$^{2}${\ensuremath{\Pi}}$_{r}$) in
  IRC +10216: A New Interstellar Phosphorus-containing Species}.
\newblock \emph{\apjl} 677, L101.
\newblock \doi{10.1086/588024}
\bibAnnoteFile{halfen08}

\bibitem[{{Helminger} and {Gordy}(1969)}]{heg69}
{Helminger}, P. and {Gordy}, W. (1969).
\newblock {Submillimeter-Wave Spectra of Ammonia and Phosphine}.
\newblock \emph{Physical Review} 188, 100--108.
\newblock \doi{10.1103/PhysRev.188.100}
\bibAnnoteFile{heg69}

\bibitem[{{Hoeft} et~al.(1972){Hoeft}, {Tiemann}, and {T{\"o}rring}}]{hoeft72}
{Hoeft}, J., {Tiemann}, E., and {T{\"o}rring}, T. (1972).
\newblock {Rotationsspektrum des PN}.
\newblock \emph{Zeitschrift Naturforschung Teil A} 27, 703--704.
\newblock \doi{10.1515/zna-1972-0424}
\bibAnnoteFile{hoeft72}

\bibitem[{{Jenkins}(2009)}]{jenkins09}
{Jenkins}, E.~B. (2009).
\newblock {A Unified Representation of Gas-Phase Element Depletions in the
  Interstellar Medium}.
\newblock \emph{\apj} 700, 1299--1348.
\newblock \doi{10.1088/0004-637X/700/2/1299}
\bibAnnoteFile{jenkins09}

\bibitem[{{Jenkins} et~al.(1986){Jenkins}, {Savage}, and {Spitzer}}]{jenkins86}
{Jenkins}, E.~B., {Savage}, B.~D., and {Spitzer}, J., L. (1986).
\newblock {Abundances of Interstellar Atoms from Ultraviolet Absorption Lines}.
\newblock \emph{\apj} 301, 355.
\newblock \doi{10.1086/163906}
\bibAnnoteFile{jenkins86}

\bibitem[{{Jim{\'e}nez-Serra} et~al.(2018){Jim{\'e}nez-Serra}, {Viti},
  {Qu{\'e}nard}, and {Holdship}}]{jimenez18}
{Jim{\'e}nez-Serra}, I., {Viti}, S., {Qu{\'e}nard}, D., and {Holdship}, J.
  (2018).
\newblock {The Chemistry of Phosphorus-bearing Molecules under Energetic
  Phenomena}.
\newblock \emph{\apj} 862, 128.
\newblock \doi{10.3847/1538-4357/aacdf2}
\bibAnnoteFile{jimenez18}

\bibitem[{{Johns} et~al.(1971){Johns}, {Stone}, and {Winnewisser}}]{johns71}
{Johns}, J.~W.~C., {Stone}, J.~M.~R., and {Winnewisser}, G. (1971).
\newblock {Millimeter wave spectra of HCP and DCP}.
\newblock \emph{Journal of Molecular Spectroscopy} 38, 437--440.
\newblock \doi{10.1016/0022-2852(71)90127-5}
\bibAnnoteFile{johns71}

\bibitem[{{J{\o}rgensen} et~al.(2020){J{\o}rgensen}, {Belloche}, and
  {Garrod}}]{jorgensen20}
{J{\o}rgensen}, J.~K., {Belloche}, A., and {Garrod}, R.~T. (2020).
\newblock {Astrochemistry During the Formation of Stars}.
\newblock \emph{\araa} 58, 727--778.
\newblock \doi{10.1146/annurev-astro-032620-021927}
\bibAnnoteFile{jorgensen20}

\bibitem[{{Jura} and {York}(1978)}]{jey78}
{Jura}, M. and {York}, D.~G. (1978).
\newblock {Observations of interstellar chlorine and phosphorus.}
\newblock \emph{\apj} 219, 861--869.
\newblock \doi{10.1086/155847}
\bibAnnoteFile{jey78}

\bibitem[{{Kawaguchi} et~al.(1983){Kawaguchi}, {Saito}, and
  {Hirota}}]{kawaguchi83}
{Kawaguchi}, K., {Saito}, S., and {Hirota}, E. (1983).
\newblock {Far-infrared laser magnetic resonance detection and microwave
  spectroscopy of the PO radical}.
\newblock \emph{\jcp} 79, 629--634.
\newblock \doi{10.1063/1.445810}
\bibAnnoteFile{kawaguchi83}

\bibitem[{{Kaye} and {Strobel}(1983)}]{kes83}
{Kaye}, J.~A. and {Strobel}, D.~F. (1983).
\newblock {Phosphine photochemistry in Saturn{\textquoteright}s atmosphere}.
\newblock \emph{\grl} 10, 957--960.
\newblock \doi{10.1029/GL010i010p00957}
\bibAnnoteFile{kes83}

\bibitem[{{Koelemay} et~al.(2022){Koelemay}, {Burton}, {Singh}, {Sheridan},
  {Bernal}, and {Ziurys}}]{koelemay22}
{Koelemay}, L.~A., {Burton}, M.~A., {Singh}, A.~P., {Sheridan}, P.~M.,
  {Bernal}, J.~J., and {Ziurys}, L.~M. (2022).
\newblock {Laboratory and Astronomical Detection of the SiP Radical
  (X$^{2}${\ensuremath{\Pi}}$_{ i }$): More Circumstellar Phosphorus}.
\newblock \emph{\apjl} 940, L11.
\newblock \doi{10.3847/2041-8213/ac9d9b}
\bibAnnoteFile{koelemay22}

\bibitem[{{Koelemay} et~al.(2023){Koelemay}, {Gold}, and {Ziurys}}]{koelemay23}
{Koelemay}, L.~A., {Gold}, K.~R., and {Ziurys}, L.~M. (2023).
\newblock {Phosphorus-bearing molecules PO and PN at the edge of the Galaxy}.
\newblock \emph{\nat} 623, 292--295.
\newblock \doi{10.1038/s41586-023-06616-1}
\bibAnnoteFile{koelemay23}

\bibitem[{{Koo} et~al.(2013){Koo}, {Lee}, {Moon}, {Yoon}, and
  {Raymond}}]{koo13}
{Koo}, B.-C., {Lee}, Y.-H., {Moon}, D.-S., {Yoon}, S.-C., and {Raymond}, J.~C.
  (2013).
\newblock {Phosphorus in the Young Supernova Remnant Cassiopeia A}.
\newblock \emph{Science} 342, 1346--1348.
\newblock \doi{10.1126/science.1243823}
\bibAnnoteFile{koo13}

\bibitem[{{Larson} et~al.(1977){Larson}, {Treffers}, and {Fink}}]{larson77}
{Larson}, H.~P., {Treffers}, R.~R., and {Fink}, U. (1977).
\newblock {Phosphine in Jupiter's atmosphere: the evidence from high-altitude
  observations at 5 micrometers.}
\newblock \emph{\apj} 211, 972--979.
\newblock \doi{10.1086/155009}
\bibAnnoteFile{larson77}

\bibitem[{{Lebouteiller} et~al.(2013){Lebouteiller}, {Heap}, {Hubeny}, and
  {Kunth}}]{lebouteiller13}
{Lebouteiller}, V., {Heap}, S., {Hubeny}, I., and {Kunth}, D. (2013).
\newblock {Chemical enrichment and physical conditions in I Zw 18}.
\newblock \emph{\aap} 553, A16.
\newblock \doi{10.1051/0004-6361/201220948}
\bibAnnoteFile{lebouteiller13}

\bibitem[{{Lebouteiller} et~al.(2006{\natexlab{a}}){Lebouteiller}, {Kuassivi},
  and {Ferlet}}]{lebouteiller06}
{Lebouteiller}, V., {Kuassivi}, and {Ferlet}, R. (2006{\natexlab{a}}).
\newblock {Phosphorus in the diffuse ISM}.
\newblock In \emph{Astrophysics in the Far Ultraviolet: Five Years of Discovery
  with FUSE}, eds. G.~{Sonneborn}, H.~W. {Moos}, and B.~G. {Andersson}. vol.
  348 of \emph{Astronomical Society of the Pacific Conference Series}, 480
\bibAnnoteFile{lebouteiller06}

\bibitem[{{Lebouteiller} et~al.(2006{\natexlab{b}}){Lebouteiller}, {Kunth},
  {Lequeux}, {Aloisi}, {D{\'e}sert}, {H{\'e}brard} et~al.}]{lebouteiller06b}
{Lebouteiller}, V., {Kunth}, D., {Lequeux}, J., {Aloisi}, A., {D{\'e}sert},
  J.~M., {H{\'e}brard}, G., et~al. (2006{\natexlab{b}}).
\newblock {Interstellar abundances in the neutral and ionized gas of NGC 604}.
\newblock \emph{\aap} 459, 161--174.
\newblock \doi{10.1051/0004-6361:20053161}
\bibAnnoteFile{lebouteiller06b}

\bibitem[{{Lefloch} et~al.(2016){Lefloch}, {Vastel}, {Viti}, {Jimenez-Serra},
  {Codella}, {Podio} et~al.}]{lefloch16}
{Lefloch}, B., {Vastel}, C., {Viti}, S., {Jimenez-Serra}, I., {Codella}, C.,
  {Podio}, L., et~al. (2016).
\newblock {Phosphorus-bearing molecules in solar-type star-forming regions:
  first PO detection}.
\newblock \emph{\mnras} 462, 3937--3944.
\newblock \doi{10.1093/mnras/stw1918}
\bibAnnoteFile{lefloch16}

\bibitem[{{Lodders}(2003)}]{lodders03}
{Lodders}, K. (2003).
\newblock {Solar System Abundances and Condensation Temperatures of the
  Elements}.
\newblock \emph{\apj} 591, 1220--1247.
\newblock \doi{10.1086/375492}
\bibAnnoteFile{lodders03}

\bibitem[{{Maas} et~al.(2019){Maas}, {Cescutti}, and {Pilachowski}}]{maas19}
{Maas}, Z.~G., {Cescutti}, G., and {Pilachowski}, C.~A. (2019).
\newblock {Phosphorus Abundances in the Hyades and Galactic Disk}.
\newblock \emph{\aj} 158, 219.
\newblock \doi{10.3847/1538-3881/ab4a1a}
\bibAnnoteFile{maas19}

\bibitem[{{Macia'}(2005)}]{macia05}
{Macia'}, E. (2005).
\newblock {The role of phosphorus in chemical evolution}.
\newblock \emph{Chem. Soc. Rev.} 34, 691--701
\bibAnnoteFile{macia05}

\bibitem[{{MacKay} and {Charnley}(2001)}]{mec01}
{MacKay}, D.~D.~S. and {Charnley}, S.~B. (2001).
\newblock {Phosphorus in circumstellar envelopes}.
\newblock \emph{\mnras} 325, 545--549.
\newblock \doi{10.1046/j.1365-8711.2001.04429.x}
\bibAnnoteFile{mec01}

\bibitem[{{Maki} and {Lovas}(1981)}]{mel81}
{Maki}, A.~G. and {Lovas}, F.~J. (1981).
\newblock {The infrared spectrum of $^{31}$P $^{14}$N near 1300 cm $^{-1}$}.
\newblock \emph{Journal of Molecular Spectroscopy} 85, 368--374.
\newblock \doi{10.1016/0022-2852(81)90209-5}
\bibAnnoteFile{mel81}

\bibitem[{{Marks} et~al.(2012){Marks}, {Kroupa}, {Dabringhausen}, and
  {Pawlowski}}]{marks12}
{Marks}, M., {Kroupa}, P., {Dabringhausen}, J., and {Pawlowski}, M.~S. (2012).
\newblock {Evidence for top-heavy stellar initial mass functions with
  increasing density and decreasing metallicity}.
\newblock \emph{\mnras} 422, 2246--2254.
\newblock \doi{10.1111/j.1365-2966.2012.20767.x}
\bibAnnoteFile{marks12}

\bibitem[{{Mart{\'\i}n} et~al.(2021){Mart{\'\i}n}, {Mangum}, {Harada},
  {Costagliola}, {Sakamoto}, {Muller} et~al.}]{martin21}
{Mart{\'\i}n}, S., {Mangum}, J.~G., {Harada}, N., {Costagliola}, F.,
  {Sakamoto}, K., {Muller}, S., et~al. (2021).
\newblock {ALCHEMI, an ALMA Comprehensive High-resolution Extragalactic
  Molecular Inventory. Survey presentation and first results from the ACA
  array}.
\newblock \emph{\aap} 656, A46.
\newblock \doi{10.1051/0004-6361/202141567}
\bibAnnoteFile{martin21}

\bibitem[{{Matthews} et~al.(1987){Matthews}, {Feldman}, and
  {Bernath}}]{matthews87}
{Matthews}, H.~E., {Feldman}, P.~A., and {Bernath}, P.~F. (1987).
\newblock {Upper Limits to Interstellar PO}.
\newblock \emph{\apj} 312, 358.
\newblock \doi{10.1086/164881}
\bibAnnoteFile{matthews87}

\bibitem[{{McElroy} et~al.(2013){McElroy}, {Walsh}, {Markwick}, {Cordiner},
  {Smith}, and {Millar}}]{mcelroy13}
{McElroy}, D., {Walsh}, C., {Markwick}, A.~J., {Cordiner}, M.~A., {Smith}, K.,
  and {Millar}, T.~J. (2013).
\newblock {The UMIST database for astrochemistry 2012}.
\newblock \emph{\aap} 550, A36.
\newblock \doi{10.1051/0004-6361/201220465}
\bibAnnoteFile{mcelroy13}

\bibitem[{{McGuire}(2022)}]{mcguire22}
{McGuire}, B.~A. (2022).
\newblock {2021 Census of Interstellar, Circumstellar, Extragalactic,
  Protoplanetary Disk, and Exoplanetary Molecules}.
\newblock \emph{\apjs} 259, 30.
\newblock \doi{10.3847/1538-4365/ac2a48}
\bibAnnoteFile{mcguire22}

\bibitem[{{Milam} et~al.(2008){Milam}, {Halfen}, {Tenenbaum}, {Apponi},
  {Woolf}, and {Ziurys}}]{milam08}
{Milam}, S.~N., {Halfen}, D.~T., {Tenenbaum}, E.~D., {Apponi}, A.~J., {Woolf},
  N.~J., and {Ziurys}, L.~M. (2008).
\newblock {Constraining Phosphorus Chemistry in Carbon- and Oxygen-Rich
  Circumstellar Envelopes: Observations of PN, HCP, and CP}.
\newblock \emph{\apj} 684, 618--625.
\newblock \doi{10.1086/589135}
\bibAnnoteFile{milam08}

\bibitem[{{Millar}(1991)}]{millar91}
{Millar}, T.~J. (1991).
\newblock {Phosphorus chemistry in dense interstellar clouds.}
\newblock \emph{\aap} 242, 241
\bibAnnoteFile{millar91}

\bibitem[{{Millar} et~al.(1987){Millar}, {Bennett}, and {Herbst}}]{millar87}
{Millar}, T.~J., {Bennett}, A., and {Herbst}, E. (1987).
\newblock {An efficient gas phase synthesis for interstellar PN}.
\newblock \emph{\mnras} 229, 41P--44.
\newblock \doi{10.1093/mnras/229.1.41P}
\bibAnnoteFile{millar87}

\bibitem[{{Mininni} et~al.(2018){Mininni}, {Fontani}, {Rivilla}, {Beltr{\'a}n},
  {Caselli}, and {Vasyunin}}]{mininni18}
{Mininni}, C., {Fontani}, F., {Rivilla}, V.~M., {Beltr{\'a}n}, M.~T.,
  {Caselli}, P., and {Vasyunin}, A. (2018).
\newblock {On the origin of phosphorus nitride in star-forming regions}.
\newblock \emph{\mnras} 476, L39--L44.
\newblock \doi{10.1093/mnrasl/sly026}
\bibAnnoteFile{mininni18}

\bibitem[{{Molpeceres} and {K{\"a}stner}(2021)}]{mek21}
{Molpeceres}, G. and {K{\"a}stner}, J. (2021).
\newblock {Computational Study of the Hydrogenation Sequence of the Phosphorous
  Atom on Interstellar Dust Grains}.
\newblock \emph{\apj} 910, 55.
\newblock \doi{10.3847/1538-4357/abe38c}
\bibAnnoteFile{mek21}

\bibitem[{{Molpeceres} et~al.(2023){Molpeceres}, {Zaverkin}, {Furuya},
  {Aikawa}, and {K{\"a}stner}}]{molpeceres23}
{Molpeceres}, G., {Zaverkin}, V., {Furuya}, K., {Aikawa}, Y., and
  {K{\"a}stner}, J. (2023).
\newblock {Reaction dynamics on amorphous solid water surfaces using
  interatomic machine-learned potentials. Microscopic energy partition revealed
  from the P + H {\textrightarrow} PH reaction}.
\newblock \emph{\aap} 673, A51.
\newblock \doi{10.1051/0004-6361/202346073}
\bibAnnoteFile{molpeceres23}

\bibitem[{{M{\"u}ller}(2013)}]{muller13}
{M{\"u}ller}, H. S.~P. (2013).
\newblock {Spectroscopic parameters of phosphine, PH$_{3}$, in its ground
  vibrational state}.
\newblock \emph{\jqsrt} 130, 335--340.
\newblock \doi{10.1016/j.jqsrt.2013.05.002}
\bibAnnoteFile{muller13}

\bibitem[{{Nandakumar} et~al.(2022){Nandakumar}, {Ryde}, {Montelius},
  {Thorsbro}, {J{\"o}nsson}, and {Mace}}]{nandakumar22}
{Nandakumar}, G., {Ryde}, N., {Montelius}, M., {Thorsbro}, B., {J{\"o}nsson},
  H., and {Mace}, G. (2022).
\newblock {The Galactic chemical evolution of phosphorus observed with IGRINS}.
\newblock \emph{\aap} 668, A88.
\newblock \doi{10.1051/0004-6361/202244724}
\bibAnnoteFile{nandakumar22}

\bibitem[{{Padovani} et~al.(2009){Padovani}, {Galli}, and
  {Glassgold}}]{padovani09}
{Padovani}, M., {Galli}, D., and {Glassgold}, A.~E. (2009).
\newblock {Cosmic-ray ionization of molecular clouds}.
\newblock \emph{\aap} 501, 619--631.
\newblock \doi{10.1051/0004-6361/200911794}
\bibAnnoteFile{padovani09}

\bibitem[{{Pantaleone} et~al.(2021){Pantaleone}, {Corno}, {Rimola}, {Balucani},
  and {Ugliengo}}]{pantaleone21}
{Pantaleone}, S., {Corno}, M., {Rimola}, A., {Balucani}, N., and {Ugliengo}, P.
  (2021).
\newblock {Ab Initio Computational Study on Fe2NiP Schreibersite: Bulk and
  Surface Characterization}.
\newblock \emph{ACS Earth and Space Chemistry} 5, 1741--1751.
\newblock \doi{10.1021/acsearthspacechem.1c00083}
\bibAnnoteFile{pantaleone21}

\bibitem[{{Pantaleone} et~al.(2023){Pantaleone}, {Corno}, {Rimola}, {Balucani},
  and {Ugliengo}}]{pantaleone23}
{Pantaleone}, S., {Corno}, M., {Rimola}, A., {Balucani}, N., and {Ugliengo}, P.
  (2023).
\newblock {Computational Study on the Water Corrosion Process at Schreibersite
  (Fe2NiP) Surfaces: from Phosphide to Phosphates}.
\newblock \emph{ACS Earth and Space Chemistry} 7, 2050--2061.
\newblock \doi{10.1021/acsearthspacechem.3c00167}
\bibAnnoteFile{pantaleone23}

\bibitem[{{Pasek} and {Lauretta}(2005)}]{pel05}
{Pasek}, M.~A. and {Lauretta}, D.~S. (2005).
\newblock {Aqueous Corrosion of Phosphide Minerals from Iron Meteorites: A
  Highly Reactive Source of Prebiotic Phosphorus on the Surface of the Early
  Earth}.
\newblock \emph{Astrobiology} 5, 515--535.
\newblock \doi{10.1089/ast.2005.5.515}
\bibAnnoteFile{pel05}

\bibitem[{{Pickett} et~al.(1998){Pickett}, {Poynter}, {Cohen}, {Delitsky},
  {Pearson}, and {M{\"u}ller}}]{pickett98}
{Pickett}, H.~M., {Poynter}, R.~L., {Cohen}, E.~A., {Delitsky}, M.~L.,
  {Pearson}, J.~C., and {M{\"u}ller}, H.~S.~P. (1998).
\newblock {Submillimeter, millimeter and microwave spectral line catalog.}
\newblock \emph{\jqsrt} 60, 883--890.
\newblock \doi{10.1016/S0022-4073(98)00091-0}
\bibAnnoteFile{pickett98}

\bibitem[{{Pirronello} et~al.(1999){Pirronello}, {Liu}, {Roser}, and
  {Vidali}}]{pirronello99}
{Pirronello}, V., {Liu}, C., {Roser}, J.~E., and {Vidali}, G. (1999).
\newblock {Measurements of molecular hydrogen formation on carbonaceous
  grains}.
\newblock \emph{\aap} 344, 681--686
\bibAnnoteFile{pirronello99}

\bibitem[{{Plane} et~al.(2021){Plane}, {Feng}, and {Douglas}}]{plane21}
{Plane}, J., {Feng}, W., and {Douglas}, K. (2021).
\newblock {Phosphorus Chemistry in Planetary Upper Atmospheres}.
\newblock In \emph{European Planetary Science Congress}. EPSC2021--322.
\newblock \doi{10.5194/epsc2021-322}
\bibAnnoteFile{plane21}

\bibitem[{{Qin} et~al.(2023){Qin}, {Hu}, {Li}, and {Liu}}]{qin23}
{Qin}, Z., {Hu}, P., {Li}, J., and {Liu}, L. (2023).
\newblock {Radiative association of P$^{+}$($^{3}$P) and O($^{3}$P) for the
  PO$^{+}$ formation}.
\newblock \emph{\mnras} 523, 2684--2692.
\newblock \doi{10.1093/mnras/stad1571}
\bibAnnoteFile{qin23}

\bibitem[{{Ranasinghe} and {Leahy}(2022)}]{ranasinghe22}
{Ranasinghe}, S. and {Leahy}, D. (2022).
\newblock {Distances, Radial Distribution, and Total Number of Galactic
  Supernova Remnants}.
\newblock \emph{\apj} 940, 63.
\newblock \doi{10.3847/1538-4357/ac940a}
\bibAnnoteFile{ranasinghe22}

\bibitem[{{Ritchey} et~al.(2023){Ritchey}, {Brown}, {Federman}, and
  {Sonnentrucker}}]{ritchey23}
{Ritchey}, A.~M., {Brown}, J.~M., {Federman}, S.~R., and {Sonnentrucker}, P.
  (2023).
\newblock {A Reexamination of Phosphorus and Chlorine Depletions in the Diffuse
  Interstellar Medium}.
\newblock \emph{\apj} 948, 139.
\newblock \doi{10.3847/1538-4357/acc179}
\bibAnnoteFile{ritchey23}

\bibitem[{{Ritchey} et~al.(2018){Ritchey}, {Federman}, and
  {Lambert}}]{ritchey18}
{Ritchey}, A.~M., {Federman}, S.~R., and {Lambert}, D.~L. (2018).
\newblock {Abundances and Depletions of Neutron-capture Elements in the
  Interstellar Medium}.
\newblock \emph{\apjs} 236, 36.
\newblock \doi{10.3847/1538-4365/aab71e}
\bibAnnoteFile{ritchey18}

\bibitem[{{Rivilla} et~al.(2020){Rivilla}, {Drozdovskaya}, {Altwegg},
  {Caselli}, {Beltr{\'a}n}, {Fontani} et~al.}]{rivilla20}
{Rivilla}, V.~M., {Drozdovskaya}, M.~N., {Altwegg}, K., {Caselli}, P.,
  {Beltr{\'a}n}, M.~T., {Fontani}, F., et~al. (2020).
\newblock {ALMA and ROSINA detections of phosphorus-bearing molecules: the
  interstellar thread between star-forming regions and comets}.
\newblock \emph{\mnras} 492, 1180--1198.
\newblock \doi{10.1093/mnras/stz3336}
\bibAnnoteFile{rivilla20}

\bibitem[{{Rivilla} et~al.(2016){Rivilla}, {Fontani}, {Beltr{\'a}n},
  {Vasyunin}, {Caselli}, {Mart{\'\i}n-Pintado} et~al.}]{rivilla16}
{Rivilla}, V.~M., {Fontani}, F., {Beltr{\'a}n}, M.~T., {Vasyunin}, A.,
  {Caselli}, P., {Mart{\'\i}n-Pintado}, J., et~al. (2016).
\newblock {The First Detections of the Key Prebiotic Molecule PO in
  Star-forming Regions}.
\newblock \emph{\apj} 826, 161.
\newblock \doi{10.3847/0004-637X/826/2/161}
\bibAnnoteFile{rivilla16}

\bibitem[{{Rivilla} et~al.(2022){Rivilla}, {Garc{\'\i}a De La Concepci{\'o}n},
  {Jim{\'e}nez-Serra}, {Mart{\'\i}n-Pintado}, {Colzi}, {Tercero}
  et~al.}]{rivilla22}
{Rivilla}, V.~M., {Garc{\'\i}a De La Concepci{\'o}n}, J., {Jim{\'e}nez-Serra},
  I., {Mart{\'\i}n-Pintado}, J., {Colzi}, L., {Tercero}, B., et~al. (2022).
\newblock {Ionize Hard: Interstellar PO+ Detection}.
\newblock \emph{Frontiers in Astronomy and Space Sciences} 9, 829288.
\newblock \doi{10.3389/fspas.2022.829288}
\bibAnnoteFile{rivilla22}

\bibitem[{{Rivilla} et~al.(2018){Rivilla}, {Jim{\'e}nez-Serra}, {Zeng},
  {Mart{\'\i}n}, {Mart{\'\i}n-Pintado}, {Armijos-Abenda{\~n}o}
  et~al.}]{rivilla18}
{Rivilla}, V.~M., {Jim{\'e}nez-Serra}, I., {Zeng}, S., {Mart{\'\i}n}, S.,
  {Mart{\'\i}n-Pintado}, J., {Armijos-Abenda{\~n}o}, J., et~al. (2018).
\newblock {Phosphorus-bearing molecules in the Galactic Center}.
\newblock \emph{\mnras} 475, L30--L34.
\newblock \doi{10.1093/mnrasl/slx208}
\bibAnnoteFile{rivilla18}

\bibitem[{{Roederer} et~al.(2014){Roederer}, {Jacobson}, {Thanathibodee},
  {Frebel}, and {Toller}}]{roederer14}
{Roederer}, I.~U., {Jacobson}, H.~R., {Thanathibodee}, T., {Frebel}, A., and
  {Toller}, E. (2014).
\newblock {Detection of Neutral Phosphorus in the Near-ultraviolet Spectra of
  Late-type Stars}.
\newblock \emph{\apj} 797, 69.
\newblock \doi{10.1088/0004-637X/797/1/69}
\bibAnnoteFile{roederer14}

\bibitem[{{Saito} et~al.(1989){Saito}, {Yamamoto}, {Kawaguchi}, {Ohishi},
  {Suzuki}, {Ishikawa} et~al.}]{saito89}
{Saito}, S., {Yamamoto}, S., {Kawaguchi}, K., {Ohishi}, M., {Suzuki}, H.,
  {Ishikawa}, S.-I., et~al. (1989).
\newblock {The Microwave Spectrum of the CP Radical and Related Astronomical
  Search}.
\newblock \emph{\apj} 341, 1114.
\newblock \doi{10.1086/167570}
\bibAnnoteFile{saito89}

\bibitem[{{Savage} and {Sembach}(1996)}]{ses96}
{Savage}, B.~D. and {Sembach}, K.~R. (1996).
\newblock {Interstellar Gas-Phase Abundances and Physical Conditions toward Two
  Distant High-Latitude Halo Stars}.
\newblock \emph{\apj} 470, 893.
\newblock \doi{10.1086/177919}
\bibAnnoteFile{ses96}

\bibitem[{{Schwartz}(2006)}]{schwartz06}
{Schwartz}, A.~W. (2006).
\newblock {Phosphorus in prebiotic chemistry}.
\newblock \emph{{Philosophical Transactions of the Royal Society B: Biological
  Sciences}} 361, 1743--1749
\bibAnnoteFile{schwartz06}

\bibitem[{{Shimonishi} et~al.(2021){Shimonishi}, {Izumi}, {Furuya}, and
  {Yasui}}]{shimonishi21}
{Shimonishi}, T., {Izumi}, N., {Furuya}, K., and {Yasui}, C. (2021).
\newblock {The Detection of a Hot Molecular Core in the Extreme Outer Galaxy}.
\newblock \emph{\apj} 922, 206.
\newblock \doi{10.3847/1538-4357/ac289b}
\bibAnnoteFile{shimonishi21}

\bibitem[{{Sil} et~al.(2021){Sil}, {Srivastav}, {Bhat}, {Mondal}, {Gorai},
  {Ghosh} et~al.}]{sil21}
{Sil}, M., {Srivastav}, S., {Bhat}, B., {Mondal}, S.~K., {Gorai}, P., {Ghosh},
  R., et~al. (2021).
\newblock {Chemical Complexity of Phosphorous-bearing Species in Various
  Regions of the Interstellar Medium}.
\newblock \emph{\aj} 162, 119.
\newblock \doi{10.3847/1538-3881/ac09f9}
\bibAnnoteFile{sil21}

\bibitem[{{Sousa-Silva} et~al.(2020){Sousa-Silva}, {Seager}, {Ranjan},
  {Petkowski}, {Zhan}, {Hu} et~al.}]{sousa20}
{Sousa-Silva}, C., {Seager}, S., {Ranjan}, S., {Petkowski}, J.~J., {Zhan}, Z.,
  {Hu}, R., et~al. (2020).
\newblock {Phosphine as a Biosignature Gas in Exoplanet Atmospheres}.
\newblock \emph{Astrobiology} 20, 235--268.
\newblock \doi{10.1089/ast.2018.1954}
\bibAnnoteFile{sousa20}

\bibitem[{{Sousa-Silva} et~al.(2013){Sousa-Silva}, {Yurchenko}, and
  {Tennyson}}]{sousa13}
{Sousa-Silva}, C., {Yurchenko}, S.~N., and {Tennyson}, J. (2013).
\newblock {A computed room temperature line list for phosphine}.
\newblock \emph{Journal of Molecular Spectroscopy} 288, 28--37.
\newblock \doi{10.1016/j.jms.2013.04.002}
\bibAnnoteFile{sousa13}

\bibitem[{{Souza} et~al.(2021){Souza}, {Silva}, and {Galv{\~a}o}}]{souza21}
{Souza}, A.~C., {Silva}, M.~X., and {Galv{\~a}o}, B. R.~L. (2021).
\newblock {Interconversion mechanisms of PN and PO in the interstellar medium
  through simple atom-diatom collisions}.
\newblock \emph{\mnras} 507, 1899--1903.
\newblock \doi{10.1093/mnras/stab2255}
\bibAnnoteFile{souza21}

\bibitem[{{Tafalla} and {Bachiller}(1995)}]{teb95}
{Tafalla}, M. and {Bachiller}, R. (1995).
\newblock {Ammonia Emission from Bow Shocks in the L1157 Outflow}.
\newblock \emph{\apjl} 443, L37.
\newblock \doi{10.1086/187830}
\bibAnnoteFile{teb95}

\bibitem[{{Tenenbaum} et~al.(2007){Tenenbaum}, {Woolf}, and
  {Ziurys}}]{tenenbaum07}
{Tenenbaum}, E.~D., {Woolf}, N.~J., and {Ziurys}, L.~M. (2007).
\newblock {Identification of Phosphorus Monoxide
  (X$^{2}${\ensuremath{\Pi}}$_{r}$) in VY Canis Majoris: Detection of the First
  PO Bond in Space}.
\newblock \emph{\apjl} 666, L29--L32.
\newblock \doi{10.1086/521361}
\bibAnnoteFile{tenenbaum07}

\bibitem[{{Tenenbaum} and {Ziurys}(2008)}]{tez08}
{Tenenbaum}, E.~D. and {Ziurys}, L.~M. (2008).
\newblock {A Search for Phosphine in Circumstellar Envelopes: PH$_{3}$ in IRC
  +10216 and CRL 2688?}
\newblock \emph{\apjl} 680, L121.
\newblock \doi{10.1086/589973}
\bibAnnoteFile{tez08}

\bibitem[{{Thorne} et~al.(1984){Thorne}, {Anicich}, {Prasad}, and
  {Huntress}}]{thorne84}
{Thorne}, L.~R., {Anicich}, V.~G., {Prasad}, S.~S., and {Huntress}, J., W.~T.
  (1984).
\newblock {The chemistry of phosphorus in dense interstellar clouds.}
\newblock \emph{\apj} 280, 139--143.
\newblock \doi{10.1086/161977}
\bibAnnoteFile{thorne84}

\bibitem[{{Townes} and {Schawlow}(1975)}]{tes75}
{Townes}, C.~H. and {Schawlow}, A.~L. (1975).
\newblock \emph{{Microwave spectroscopy.}} (Dover Publications)
\bibAnnoteFile{tes75}

\bibitem[{{Turner} et~al.(2018){Turner}, {Abplanalp}, {Blair}, {Dayuha}, and
  {Kaiser}}]{turner18}
{Turner}, A.~M., {Abplanalp}, M.~J., {Blair}, T.~J., {Dayuha}, R., and
  {Kaiser}, R.~I. (2018).
\newblock {An Infrared Spectroscopic Study Toward the Formation of
  Alkylphosphonic Acids and Their Precursors in Extraterrestrial Environments}.
\newblock \emph{\apjs} 234, 6.
\newblock \doi{10.3847/1538-4365/aa9183}
\bibAnnoteFile{turner18}

\bibitem[{{Turner} and {Bally}(1987)}]{teb87}
{Turner}, B.~E. and {Bally}, J. (1987).
\newblock {Detection of Interstellar PN: The First Identified Phosphorus
  Compound in the Interstellar Medium}.
\newblock \emph{\apjl} 321, L75.
\newblock \doi{10.1086/185009}
\bibAnnoteFile{teb87}

\bibitem[{{Turner} et~al.(1990){Turner}, {Tsuji}, {Bally}, {Guelin}, and
  {Cernicharo}}]{turner90}
{Turner}, B.~E., {Tsuji}, T., {Bally}, J., {Guelin}, M., and {Cernicharo}, J.
  (1990).
\newblock {Phosphorus in the Dense Interstellar Medium}.
\newblock \emph{\apj} 365, 569.
\newblock \doi{10.1086/169511}
\bibAnnoteFile{turner90}

\bibitem[{{Vasyunin} and {Herbst}(2013)}]{veh13}
{Vasyunin}, A.~I. and {Herbst}, E. (2013).
\newblock {A Unified Monte Carlo Treatment of Gas-Grain Chemistry for Large
  Reaction Networks. II. A Multiphase Gas-surface-layered Bulk Model}.
\newblock \emph{\apj} 762, 86.
\newblock \doi{10.1088/0004-637X/762/2/86}
\bibAnnoteFile{veh13}

\bibitem[{{Villanueva} et~al.(2021){Villanueva}, {Cordiner}, {Irwin}, {de
  Pater}, {Butler}, {Gurwell} et~al.}]{villanueva21}
{Villanueva}, G.~L., {Cordiner}, M., {Irwin}, P.~G.~J., {de Pater}, I.,
  {Butler}, B., {Gurwell}, M., et~al. (2021).
\newblock {No evidence of phosphine in the atmosphere of Venus from independent
  analyses}.
\newblock \emph{Nature Astronomy} 5, 631--635.
\newblock \doi{10.1038/s41550-021-01422-z}
\bibAnnoteFile{villanueva21}

\bibitem[{{Viti} et~al.(2011){Viti}, {Jimenez-Serra}, {Yates}, {Codella},
  {Vasta}, {Caselli} et~al.}]{viti11}
{Viti}, S., {Jimenez-Serra}, I., {Yates}, J.~A., {Codella}, C., {Vasta}, M.,
  {Caselli}, P., et~al. (2011).
\newblock {L1157-B1: Water and Ammonia as Diagnostics of Shock Temperature}.
\newblock \emph{\apjl} 740, L3.
\newblock \doi{10.1088/2041-8205/740/1/L3}
\bibAnnoteFile{viti11}

\bibitem[{{Wakelam} et~al.(2017){Wakelam}, {Bron}, {Cazaux}, {Dulieu}, {Gry},
  {Guillard} et~al.}]{wakelam17}
{Wakelam}, V., {Bron}, E., {Cazaux}, S., {Dulieu}, F., {Gry}, C., {Guillard},
  P., et~al. (2017).
\newblock {H$_{2}$ formation on interstellar dust grains: The viewpoints of
  theory, experiments, models and observations}.
\newblock \emph{Molecular Astrophysics} 9, 1--36.
\newblock \doi{10.1016/j.molap.2017.11.001}
\bibAnnoteFile{wakelam17}

\bibitem[{{Wakelam} et~al.(2012){Wakelam}, {Herbst}, {Loison}, {Smith},
  {Chandrasekaran}, {Pavone} et~al.}]{wakelam12}
{Wakelam}, V., {Herbst}, E., {Loison}, J.~C., {Smith}, I.~W.~M.,
  {Chandrasekaran}, V., {Pavone}, B., et~al. (2012).
\newblock {A KInetic Database for Astrochemistry (KIDA)}.
\newblock \emph{\apjs} 199, 21.
\newblock \doi{10.1088/0067-0049/199/1/21}
\bibAnnoteFile{wakelam12}

\bibitem[{{Willacy} and {Millar}(1997)}]{wem97}
{Willacy}, K. and {Millar}, T.~J. (1997).
\newblock {Chemistry in oxygen-rich circumstellar envelopes.}
\newblock \emph{\aap} 324, 237--248
\bibAnnoteFile{wem97}

\bibitem[{{Wilson}(1999)}]{wilson99}
{Wilson}, T.~L. (1999).
\newblock {Isotopes in the interstellar medium and circumstellar envelopes}.
\newblock \emph{Reports on Progress in Physics} 62, 143--185.
\newblock \doi{10.1088/0034-4885/62/2/002}
\bibAnnoteFile{wilson99}

\bibitem[{{Wilson} et~al.(2012){Wilson}, {Rohlfs}, and
  {Huttemeister}}]{wilson12}
{Wilson}, T.~L., {Rohlfs}, K., and {Huttemeister}, S. (2012).
\newblock \emph{{Tools of Radio Astronomy, 5th edition}}
\bibAnnoteFile{wilson12}

\bibitem[{{Woosley} and {Weaver}(1995)}]{wew95}
{Woosley}, S.~E. and {Weaver}, T.~A. (1995).
\newblock {The Evolution and Explosion of Massive Stars. II. Explosive
  Hydrodynamics and Nucleosynthesis}.
\newblock \emph{\apjs} 101, 181.
\newblock \doi{10.1086/192237}
\bibAnnoteFile{wew95}

\bibitem[{{Wurmser} and {Bergner}(2022)}]{web22}
{Wurmser}, S. and {Bergner}, J.~B. (2022).
\newblock {New Detections of Phosphorus Molecules toward Solar-type
  Protostars}.
\newblock \emph{\apj} 934, 153.
\newblock \doi{10.3847/1538-4357/ac7c0e}
\bibAnnoteFile{web22}

\bibitem[{{Wyse} et~al.(1972){Wyse}, {Manson}, and {Gordy}}]{wyse72}
{Wyse}, F.~C., {Manson}, E.~L., and {Gordy}, W. (1972).
\newblock {Millimeter Wave Rotational Spectrum and Molecular Constants of
  $^{31}$P$^{14}$N}.
\newblock \emph{\jcp} 57, 1106--1108.
\newblock \doi{10.1063/1.1678365}
\bibAnnoteFile{wyse72}

\bibitem[{{Yamaguchi} et~al.(2011){Yamaguchi}, {Takano}, {Sakai}, {Sakai},
  {Liu}, {Su} et~al.}]{yamaguchi11}
{Yamaguchi}, T., {Takano}, S., {Sakai}, N., {Sakai}, T., {Liu}, S.-Y., {Su},
  Y.-N., et~al. (2011).
\newblock {Detection of Phosphorus Nitride in the Lynds 1157 B1 Shocked
  Region}.
\newblock \emph{\pasj} 63, L37--L41.
\newblock \doi{10.1093/pasj/63.5.L37}
\bibAnnoteFile{yamaguchi11}

\bibitem[{{Yamamoto}(2017)}]{yamamoto17}
{Yamamoto}, S. (2017).
\newblock \emph{{Introduction to Astrochemistry: Chemical Evolution from
  Interstellar Clouds to Star and Planet Formation}} (Springer Japan).
\newblock \doi{10.1007/978-4-431-54171-4}
\bibAnnoteFile{yamamoto17}

\bibitem[{{Ziurys}(1987)}]{ziurys87}
{Ziurys}, L.~M. (1987).
\newblock {Detection of Interstellar PN: The First Phosphorus-bearing Species
  Observed in Molecular Clouds}.
\newblock \emph{\apjl} 321, L81.
\newblock \doi{10.1086/185010}
\bibAnnoteFile{ziurys87}

\bibitem[{{Ziurys} et~al.(2007){Ziurys}, {Milam}, {Apponi}, and
  {Woolf}}]{ziurys07}
{Ziurys}, L.~M., {Milam}, S.~N., {Apponi}, A.~J., and {Woolf}, N.~J. (2007).
\newblock {Chemical complexity in the winds of the oxygen-rich supergiant star
  VY Canis Majoris}.
\newblock \emph{\nat} 447, 1094--1097.
\newblock \doi{10.1038/nature05905}
\bibAnnoteFile{ziurys07}

\bibitem[{{Ziurys} et~al.(2018){Ziurys}, {Schmidt}, and {Bernal}}]{ziurys18}
{Ziurys}, L.~M., {Schmidt}, D.~R., and {Bernal}, J.~J. (2018).
\newblock {New Circumstellar Sources of PO and PN: The Increasing Role of
  Phosphorus Chemistry in Oxygen-rich Stars}.
\newblock \emph{\apj} 856, 169.
\newblock \doi{10.3847/1538-4357/aaafc6}
\bibAnnoteFile{ziurys18}

\end{thebibliography}

\end{document}